%% file: main.tex
\documentclass[conference,compsoc]{IEEEtran}

\usepackage[nocompress,noadjust]{cite}
\usepackage[pdftex]{graphicx}
\usepackage[hyphens]{url}
\usepackage[pdftex,bookmarksnumbered,hidelinks,breaklinks]{hyperref}
\usepackage[caption=false,font=footnotesize,labelfont=sf,textfont=sf]{subfig}

\usepackage{
  booktabs,
  enumitem,
  amsmath,
  amsthm,
  amsfonts,
  framed,
  color,
  float,
  wrapfig,
  graphics,
  mathtools,
  multirow,
  nicefrac,
  makecell,
  soul,
  colortbl,
  etoolbox,
  relsize,
  anyfontsize,
}

\newenvironment{SUBENVccomment}[2]{\color{#1}[#2:]~}{\color{black}}

\definecolor{author1}{rgb}     {0.9,0.5,0.0}
\definecolor{author2}{rgb}     {0.6,0.0,0.8}
\definecolor{author3}{rgb}     {0.0,0.5,0.0}
\definecolor{author4}{rgb}     {0.9,0.2,0.2}

\newcommand{\commentout}[1]{}

\newcommand{\floor}[1]{\left \lfloor #1 \right \rfloor }

\newtheoremstyle{abcd}
  {0.1cm}
  {0.05cm}
  {\itshape}
  {}
  {\bfseries}
  {.}
  {.5em}
  {}

\theoremstyle{abcd}
\newtheorem*{mytheorem}{\textbf{\textit{Watermark Removal}}}
\BeforeBeginEnvironment{mytheorem}{\noindent\ignorespaces}

\makeatletter
\patchcmd{\@thm}
  {\trivlist}
  {\list{}{\leftmargin=\thm@leftmargin\rightmargin=\thm@rightmargin}}
  {}{}
\patchcmd{\@endtheorem}
  {\endtrivlist}
  {\endlist}
  {}{}
\newlength{\thm@leftmargin}
\newlength{\thm@rightmargin}
\newcommand{\xnewtheorem}[3]{%
  \newenvironment{#3}
    {\thm@leftmargin=#1\relax\thm@rightmargin=#2\relax\begin{#3INNER}}
    {\end{#3INNER}}%
  \newtheorem*{#3INNER}%
}

\newcommand{\thickhline}{%
    \noalign {\ifnum 0=`}\fi \hrule height 2pt 
    \futurelet \reserved@a \@xhline 
}

\newcommand{\topthickhline}{%
    \noalign {\ifnum 0=`}\fi \hrule height 1pt 
    \futurelet \reserved@a \@xhline
}

\newcolumntype{"}{@{\hskip\tabcolsep\vrule width 1pt\hskip\tabcolsep}}
\makeatother

\begin{document}

\title{UnMarker: A Universal Attack on Defensive Image Watermarking\\
{\small \textit{To appear at IEEE S\&P 2025. Author version.}}\vspace{-3mm}}

\author{\IEEEauthorblockN{Andre Kassis\IEEEauthorrefmark{1} and Urs Hengartner\IEEEauthorrefmark{2}}
\IEEEauthorblockA{\textit{Cheriton School of Computer Science} \\
\textit{University of Waterloo}\\
Waterloo, Canada \\
\IEEEauthorrefmark{1}akassis@uwaterloo.ca, \IEEEauthorrefmark{2}urs.hengartner@uwaterloo.ca}
}

\maketitle


\input{abstract}
\IEEEpeerreviewmaketitle
\input{Sections/Introduction}
\input{Sections/background}
\input{Sections/threat_model}

\input{Sections/theory}
\input{Sections/Experiments}
\input{Sections/discussion}
\input{Sections/Conclusion}
\input{Sections/ack}
\bibliographystyle{IEEEtran}
\bibliography{IEEEabrv,main}
\appendices
\input{Sections/appendix}
\input{Sections/meta_review}
\end{document}

%% file: abstract.tex
\begin{abstract}
Reports regarding the misuse of \textit{Generative AI} (\textit{GenAI}) to create deepfakes are frequent. \textit{Defensive watermarking} enables \textit{GenAI} providers to hide fingerprints in their images and use them later for deepfake detection. Yet, its potential has not been fully explored. We present \textit{UnMarker}--- the first practical \textit{universal} attack on defensive watermarking. Unlike existing attacks, \textit{UnMarker} requires no detector feedback, no unrealistic knowledge of the watermarking scheme or similar models, and no advanced denoising pipelines that may not be available. Instead, being the product of an in-depth analysis of the watermarking paradigm revealing that robust schemes must construct their watermarks in the spectral amplitudes, \textit{UnMarker} employs two novel adversarial optimizations to disrupt the spectra of watermarked images, erasing the watermarks. Evaluations against \textit{SOTA} schemes prove \textit{UnMarker's} effectiveness. It not only defeats traditional schemes while retaining superior quality compared to existing attacks but also breaks \textit{semantic} watermarks that alter an image's structure, reducing the best detection rate to $43\%$ and rendering them useless. To our knowledge, \textit{UnMarker} is the first practical attack on \textit{semantic} watermarks, which have been deemed the future of defensive watermarking. Our findings show that defensive watermarking is not a viable defense against deepfakes, and we urge the community to explore alternatives. 
\end{abstract}

%% file: Sections/Introduction.tex
\section{Introduction}
\label{sec:intro} 

\textit{GenAI} has amassed popularity due to its growing sophistication and accessibility~\cite{GoogleAI,Clipdrop,DALLE2}. This technology has solved problems in various fields~\cite{cgan,radiology,plastic,WIPO}. Yet, it comes with the risk of malicious entities using it to generate harmful images (deepfakes)~\cite{de2021distinct,social_stability}, proven by many reports of those being used in political smear campaigns~\cite{political} and non-consensual pornography~\cite{Harwell_2018}.

Defenses first emerged as passive  detectors~\cite{detector2,detector3,detector4,detector5} that tell apart fake from real images, but these were proven ineffective, failing to generalize to unseen synthesis algorithms~\cite{detector_attack3}. As training advanced generators is costly and prohibitive~\cite{costlyGen} in the absence of the extensive resources and high-quality data that only a few \textit{GenAI} providers possess, adversaries typically resort to these providers' \textit{API}s to generate harmful content. This recently led to the adoption of defensive watermarking as a countermeasure, which enables providers to embed hidden signatures in their images that can later be identified using designated detectors. Defensive watermarking has gained traction in the research community~\cite{TreeRing,HiDDeN,stablesig,yu1,yu2,ptw,stegastamp,recipe} and the industry, as providers have already implemented and are testing such solutions (e.g., \textit{Google}'s \textit{SynthID}~\cite{SynthID}) or pledged to do so soon~\cite{commitments} in commitments to the \textit{White House}~\cite{whitehouse}.

Defensive watermarking protects against limited adversaries, following the assumption that attackers lack the resources to train complex generators. Yet, existing schemes fail to demonstrate strong guarantees for this resistance beyond empirical robustness to naive manipulations. We devise \textit{UnMarker}--- the first \textit{universal} (effective against \textbf{all} schemes), \textit{black-box} (without access to the scheme's parameters or similar systems), \textit{data-free} (no additional data needed) and \textit{query-free} (without detector feedback) attack. 

Our motivating insight is that, for any scheme, there must be a \textbf{\textit{universal carrier}} that embeds the watermark in all images. In layman's terms, this \textit{carrier} is the set of all attributes the detector may examine when verifying the watermark. This set must exist and is finite (shared among all images) as detectors are computationally limited. By analyzing the desired resistance to basic manipulations and the requirement that watermarks must naturally integrate with various contents, we deduce that this \textit{carrier} can \textit{\textbf{only}} be in spectral amplitudes. That is, a ``conventionally'' robust scheme \textit{must} construct its watermarks by manipulating the frequencies at which the different pixels vary, forcing the \textit{collective} distribution of these value shifts at different rates to follow a unique pattern. We can roughly even isolate the range in which watermarks are encoded: the more robust \textit{semantic} schemes~\cite{TreeRing,stegastamp}, which alter the structure, target low frequencies while content-preserving \textit{non-semantic} watermarks~\cite{yu1,HiDDeN,ptw,stablesig} populate higher bands.

\textit{UnMarker} disrupts these spectral amplitudes. To remove \textit{non-semantic} watermarks, it \textit{directly} modifies the image to optimize a designated loss, affecting the rates at which \textit{individual} pixels vary w.r.t. their neighbors and impacting the distribution of high-frequency amplitudes. For \textit{semantic} watermarks, we devise novel optimizable filters and a procedure to learn their weights to \textit{systemically} alter the consistency at different regions, causing discrepancies at lower frequencies. Unlike known filters that rely on geometric heuristics, limiting them to minor distortions, \textit{UnMarker} is guided by a perceptual loss to maximize the disturbance in visually non-critical areas without compromising quality. As a result, the \textit{carrier} set is shifted away from the watermark's distribution. By targeting the \textit{carrier}, \textit{UnMarker} abandons the dependence on a specific model, leaving it successful without knowledge of the scheme or detector feedback. Previous adversarial attacks~\cite{saberi2023robustness,WeVade,WAVES,lukas2023leveraging,hu2024transfer,RD-IWAN} all face this limitation, failing under realistic conditions.

We experiment with seven \textit{SOTA} schemes from top venues. By combining the above techniques with \textit{mild cropping}, \textit{UnMarker} breaks them all, despite its \textit{black-box} and \textit{query-free} nature. We compare \textit{UnMarker} with the recent \textit{VAEAttack}~\cite{provable} and \textit{DiffusionAttack}~\cite{provable,saberi2023robustness,WAVES}, demonstrating far superior results, defeating \textit{semantic} watermarks against which they fail and disputing misconceptions regarding these schemes' robustness to limited adversaries~\cite{provable,WAVES,saberi2023robustness}. Our results urge vendors to rethink their response to deepfakes. We make the following contributions: 
\begin{itemize} [leftmargin=*]
\item We thoroughly analyze defensive watermarking and reveal trade-offs that make it an invalid countermeasure. 
\item We employ novel spectral optimizations to devise \textit{UnMarker}--- the first \textit{universal} \textit{data-free, black-box}, and \textit{query-free} attack against defensive watermarking\footnote{\url{https://github.com/andrekassis/ai-watermark}}.
\item We demonstrate \textit{UnMarker}'s efficacy against seven \textit{SOTA} schemes under restrictive settings, rendering them useless.
\item We compare \textit{UnMarker} to the strongest \textit{query-free black-box} attacks, attaining superior performance and quality.
\item Using a novel class of optimizable filters to incur structural disturbances, \textit{UnMarker} is the first practical attack to defeat \textit{semantic} watermarks, whose robustness has marked them as the ultimate candidates for trustworthy deepfake detection, showing this assumption is highly questionable.
\end{itemize}

%% file: Sections/background.tex
\section{Background \& Related Work}
\label{sec:Background}
\subsection{Deepfake Detection}
\label{subsec:defensive}
\noindent
\textit{\textbf{Passive Detectors.}} These systems~\cite{detector2,detector3,detector4,detector5} identify deepfakes based on semantic incoherences~\cite{incoherence1,incoherence2} and low-level artifacts~\cite{detector2,detector4,detector6,detector7}. Advanced generators can produce high-quality outputs without such telltales, and manipulations can fool these detectors~\cite{detector_attack1,detector_attack2,BreakingSCVA}. They also fail to generalize to unseen models~\cite{detector_attack3}.

\noindent
\textit{\textbf{Defensive Watermarking.}} It assumes attackers lack the resources to train high-quality generators. A watermarking scheme consists of \textit{watermark ($w$), encoder ($\mathcal{E}$)} and \textit{detector ($\mathcal{D}$)}. 
In a \textit{general-purpose} scheme~\cite{HiDDeN,stegastamp}, $\mathcal{E}$ accepts any image $x$ and outputs a similar watermarked $x_w\!\! = \!\!\mathcal{E}(x, w)$. Given image $x^\prime$, $\mathcal{D}$ extracts a sequence $w^\prime \!\!=\!\! \mathcal{D}(x^\prime)$ that will be \textit{almost} identical to $w$ $iff$ $x^\prime$ was previously watermarked. \textit{Generator-specific} approaches ``watermark'' generative models: $\mathcal{E}$ can be the generator itself trained to only produce watermarked images~\cite{yu1,yu2} or an algorithm that modifies its weights~\cite{ptw,stablesig} or latent codes~\cite{TreeRing} s.t. images it produces embed the watermark. Heuristics-based watermarks~\cite{nonlearning1,nonlearning2,invisiblewatermark} \textit{directly} manipulate spectral representations, forming a signature in select frequencies. These are removable via post-processing~\cite{WeVade,stablesig}. We focus on learning-based methods~\cite{HiDDeN,stegastamp,yu1,yu2,ptw,stablesig,TreeRing,recipe}, where $\mathcal{E}$ and $\mathcal{D}$ are \textit{DNN}s as they resist post-processing.

\noindent
\textit{\underline{\textit{Semantic} vs. \textit{Non-Semantic}:}} \textit{Non-semantic} schemes take an image and produce a visually identical watermarked version. Aside from \textit{StegaStamp}~\cite{stegastamp}, \textit{general-purpose} schemes are \textit{non-semantic}. \textit{Generator-specific} schemes, except \textit{TRW}~\cite{TreeRing}, are also \textit{non-semantic}, as their training restricts them to output images identical to non-watermarked references. \textit{Semantic} watermarks can influence the content. This term was popularized by Wen et al. with their introduction of \textit{TRW}~\cite{TreeRing} that alters the latent codes used by text-to-image \textit{Latent Diffusion Models (LDMs)}~\cite{latentdiff}. When providing a random code with a textual phrase, the \textit{LDM} uses this code as a seed to produce an image matching the text (e.g, ``a dog on a bike'' results in an image of a dog riding a bike but these objects will change based on the seed). \textit{TRW} injects a pattern in the Fourier transforms of all latent codes before generation. As a result, watermarked images will differ from non-watermarked ones but still satisfy the text. This leaves watermarks invisible but highly robust. \textit{StegaStamp}~\cite{stegastamp} precedes \textit{TRW} and operates similar to its \textit{general-purpose} counterparts. Yet, it is less stealthy, causing structural changes (see Fig~\ref{fig:example}). Still, its strategically distributed distortion ensures minimal visibility. In defensive watermarking, users cannot observe non-watermarked images, further hiding these nuances and making \textit{StegaStamp} highly applicable. While \textit{semantic watermarks} primarily refer to \textit{TRW}, previous work recognizes the similarities in both schemes, distinguishing them as more robust~\cite{provable,saberi2023robustness,WAVES}. Thus, we refer to \textit{TRW} and \textit{StegaStamp} as \textit{semantic}. 

\subsection{Attacks Against Defensive Watermarking}
\label{subsec:attacks_backround}

\noindent
\textbf{\textit{Regeneration Attacks.}}
Proposed by Zhao et al.~\cite{provable}, they rely on \textit{Diffusion Models (DMs)~\cite{DiffusionModels}} or \textit{Variational Autoencoders (VAEs)~\cite{kingma2013auto}} (henceforth, the \textit{DiffusionAttack} and \textit{VAEAttack}) to introduce noise and then reconstruct images after the watermark has been eliminated. Saberi et al.~\cite{saberi2023robustness} and \textit{WAVES}~\cite{WAVES} revisit these attacks. All these works verify \textit{non-semantic} watermarks' vulnerability, against which regeneration attacks offer theoretical guarantees. Yet, \textit{semantic} schemes remain robust. While \textit{TRW} experiences some degradation~\cite{WAVES}, it occurs at a considerable distortion where the attack is invalid (see~\S\ref{sec:Threat}). Regeneration is the only practical (\textit{black-box} and \textit{query-free}) attack but fails against \textit{semantic} watermarks. It also relies on the denoising pipelines' fidelity, often impacting quality (see~\S\ref{subsubsec:sota}). \textit{UnMarker} defeats all watermarks while also abandoning the dependence on regeneration models, ensuring quality.

\noindent
\textit{\textbf{Adversarial Attacks.}} Realistically, attackers cannot access the scheme (see~\S\ref{sec:Threat}), making traditional adversarial attacks that target the decision boundaries of specific models ineffective. Thus, existing works make relaxing assumptions, failing to retain success under practical conditions.

\textit{WEVADE}~\cite{WeVade} attacks watermarked images by querying the detector and updating them accordingly until the watermark is not detected. Yet, detector feedback may not be available (see~\S\ref{sec:Threat}), making it ineffective. Lukas et al.~\cite{lukas2023leveraging} propose \textit{query-free} attacks but assume unrealistic access to a generator with similar architecture and data (see~\S\ref{sec:Threat}). Saberi et al.~\cite{saberi2023robustness} mount an attack on a surrogate detector, which transfers to the target. Training the substitute requires non-watermarked samples from the provider's distribution, violating the defensive watermarking principles~\cite{WAVES,hu2024transfer} (see~\S\ref{sec:Threat}) as providers only release \textit{watermarked} images and possess exclusive datasets. Still, it fails against the \textit{semantic} \textit{StegaStamp} without affecting quality. \textit{RD-IWAN}~\cite{RD-IWAN} also requires access to non-watermarked provider images. An et al.~\cite{WAVES} evaluate surrogate attacks specifically on \textit{semantic} schemes, finding \textit{StegaStamp} robust, while \textit{TRW} is affected under unrealistic access to its autoencoder or images generated by it with \textit{different} watermarks, which is impractical as any provider's published image bear its \textit{single} identifying watermark. They also prove Saberi et al.'s~\cite{saberi2023robustness} surrogate attack unsuccessful absent non-watermarked provider images. Hu et al.'s~\cite{hu2024transfer} ensemble surrogate attack requires no non-watermarked provider images. Yet, the evaluations exclude \textit{semantic} watermarks and only consider surrogates and targets of the same scheme, 
leading to sufficient commonalities for transferability and violating the \textit{black-box} setting. It also requires extensive resources ($40\!\!-\!\!100$ surrogates), data, and large perturbations, making it impractical. \textit{UnMarker} waives impractical assumptions, requires no data or prohibitive resources, and preserves quality.

%% file: Sections/threat_model.tex
\section{Threat Model}
\label{sec:Threat}
Our framework considers two parties: \textbf{\textit{Providers}} who develop and monetize these platforms, and \textbf{\textit{Attackers}} abusing their access to generate malicious content.

\noindent
\textbf{\textit{Providers' Goals.}} GenAI providers must ensure their images are not used to spread disinformation. Thus, they watermark generated images. For a suspicious image, they invoke the detector to test for the watermark, proving it fake.

\noindent
\textbf{\textit{Providers' Capabilities.}} They have resources and data to train \textit{SOTA} generators. As is common practice~\cite{GoogleAI,DALLE2}, their \textit{API}s are restrictive, accepting queries (e.g., text) and returning watermarked images, hiding intermediate results.

\noindent
\textbf{\textit{Attacker's Goals.}} Attackers wish to generate real-looking deepfakes by invoking the providers' \textit{API}s and then using methods to 1) remove watermarks while 2) retaining quality.

\noindent
\textbf{\textit{Attacker's Capabilities:}} \underline{\textit{Resources.}} Following the defensive watermarking assumptions (see~\S\ref{sec:Background}), attackers lack high-quality data or extensive resources. Thus, training advanced models or surrogates for the providers' generators~\cite{lukas2023leveraging} or detectors~\cite{saberi2023robustness} is not possible. As some services (e.g., \textit{AWS}) allow renting advanced hardware, attackers may access powerful machines. Yet, this can only be short-term due to high costs, not for training large models over long \textit{GPU} hours.

\noindent
\underline{\textit{Black-Box.}} As many providers keep their algorithms and datasets secret~\cite{Google}, attackers have no whitebox access to (similar) models, data, or any information about the system. 

\noindent
\underline{\textit{Query-Free.}} When providers restrict access to their detector to trusted corporations or employees, as confirmed by recent announcements~\cite{watermarking_use}, attacks relying on detector feedback~\cite{WeVade} lose their efficacy. Given these announcements and defensive watermarking's vulnerability when query access to the detector is publicly available (see~\S\ref{subsec:attacks_backround}), we follow previous work~\cite{hu2024transfer,provable,saberi2023robustness} and investigate this paradigm against \textit{query-free} attacks where attackers cannot interact with the detector directly to observe its decisions.

%% file: Sections/theory.tex
\section{Design Philosophy}
\label{sec:Threory}
Since the detector $\mathcal{D}$ must be able to verify the watermark in \textit{\textbf{any}} image, it should know where to find this watermark, meaning there must be a set of attributes, namely the \textit{universal carrier}, that can be measured in all images to make this decision. Our analysis will reveal that identifying this set is possible since the requirements from a robust scheme eliminate all options except \textit{\textbf{the spectral amplitudes}}. Thus, any robust scheme will inject its watermarks in the collective magnitudes of the different frequency bands. In~\S\ref{sec:unmarker}, we show how identifying this \textit{carrier} enables attackers to mount destructive attacks that remove the watermark with minimal visual impact and no knowledge of the system.

\subsection{Uncovering The Universal Carrier}
\label{subsection:freq}
Any computationally limited detector $\mathcal{D}$ can only monitor a finite set of attributes $\mathcal{\boldsymbol{P}}=\{p_b|\; b\in [\![ B ]\!]\}$ in all images. Deciding if an image is watermarked relies on whether (a subset of) $\mathcal{\boldsymbol{P}}$'s attributes follow a \textit{watermark property distribution} $\mathcal{\boldsymbol{P}}_w$ that $\mathcal{D}$ associates with the watermark $w$. While $\mathcal{\boldsymbol{P}}$ is a set of attributes, $\mathcal{\boldsymbol{P}}_w$ is a distribution over $\mathcal{\boldsymbol{P}}$'s values, that identifies $w$. There may be several patterns $\mathcal{\boldsymbol{P}}_w$ pertaining to subsets of $\mathcal{\boldsymbol{P}}$. We term $\mathcal{\boldsymbol{P}}$ the \textit{universal carrier} as it contains all the properties $w$ can influence. Although $\mathcal{\boldsymbol{P}}_w$ should allow watermarking different contents, its variance must still be sufficiently low to prevent detecting $w$ in non-watermarked images, restricting it to a reasonably narrow vicinity. Thus, introducing small deviations into $\mathcal{\boldsymbol{P}}$'s values can cause any watermarked image to violate this distribution $\mathcal{\boldsymbol{P}}_w$, removing the watermark without degrading quality. This destructive strategy requires no knowledge of $\mathcal{\boldsymbol{P}}_w$, making it feasible without feedback from $\mathcal{D}$ if the \textit{carrier} $\mathcal{\boldsymbol{P}}$ is correctly identified. To identify $\mathcal{\boldsymbol{P}}$, we analyze widely-accepted requirements~\cite{TreeRing,ptw,HiDDeN,stegastamp,yu1,yu2,stablesig} from a robust scheme:

\noindent
\textit{\textbf{Robust Watermarking Requirements.}} \textit{\underline{Universal Applicabil-}} \textit{\underline{ity:}} The scheme must watermark images of various concepts and scenes, and the watermark must always be detectable.

\noindent
\textit{\underline{Universal Stealthiness.}} The process cannot introduce unnatural artifacts that degrade quality. We do not require the watermark not to leave visible traces since \textit{semantic} watermarks violate this concept by definition, influencing the structure. Yet, as users only observe watermarked images, quality becomes the only relevant metric for \textit{stealthiness}.

 \begin{figure}[t]
  \centering
\includegraphics[width=0.9\linewidth,height=3.5cm]{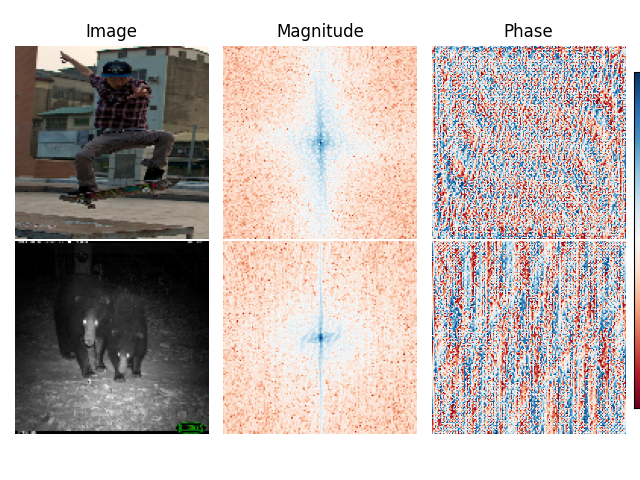}
  \caption{Spectral analysis of two images. Both depict worldly objects of high consistency. Thus, the \textit{\textbf{collective}} spectral magnitudes are distributed similarly as the low frequencies (center) corresponding to gradual variations are always far more dominant. Phases determine the spatial arrangement of the pixel value shifts that constitute these magnitudes to shape the content, making them extremely different.}
  \label{fig:spectrum}
\end{figure}

\noindent
\textit{\underline{Robustness to Manipulations.}} The watermark should resist known manipulations (e.g, \textit{cropping}, \textit{blurring}, \textit{compression}).

\subsubsection{Candidates} 
\label{subsubsec:candidates}
An image is characterized by its 1) \textit{spatial structure} (pixel values), and 2) \textit{frequency distribution}--- the rates at which values alternate between neighboring pixels. Thus, the \textit{carrier} must reside in these channels. Below, we present hypotheses, accompanied by an in-depth analysis, regarding the possibility of injecting watermarks in each of these channels, identifying a unique \textit{carrier} capable of satisfying all demands. Our claims are corroborated by \textit{UnMarker}'s empirical success in~\S\ref{sec:exp}.

\noindent
\textbf{H1. \emph{Watermarks cannot be embedded in the spatial structure (pixel values)}}. This is because they make for an \textit{unreliable carrier} that experiences sporadic variations (in the absence of a watermark) between images based on their contents (different images assign extremely different values to their pixels). However, the scheme must always bring the \textit{carrier}'s values to the same distribution $\mathcal{\boldsymbol{P}}_w$ for any watermarked image. When these values naturally (i.e., without the watermark) exhibit a high degree of content dependency, this is impossible without restricting the scheme to processing a small set of similar contents or leaving unnatural cues in many outputs corresponding to contents that would otherwise require these values in images portraying them to lie far away from the desired $\mathcal{\boldsymbol{P}}_w$. As this distance increases, the magnitudes of the changes needed to bring them to $\mathcal{\boldsymbol{P}}_w$ grow, failing to integrate naturally with the content. Restrictions violate \textit{applicability}, while unnatural artifacts are not \textit{stealthy}. The scheme may strategically place pixels encoding the watermark to accommodate large changes with minimal visibility. Two possibilities exist: either they are far apart to remain unnoticeable despite significantly differing from their environments or located in regions the content is unlikely to occupy (i.e., the background). Both are volatile and cannot withstand manipulations (e.g., \textit{cropping} or \textit{noise}), violating the \textit{robustness} requirement (see~\S\ref{subsubsec:utility}). 

 \begin{figure}[t]
  \centering
\includegraphics[width=0.9\linewidth,height=4cm]{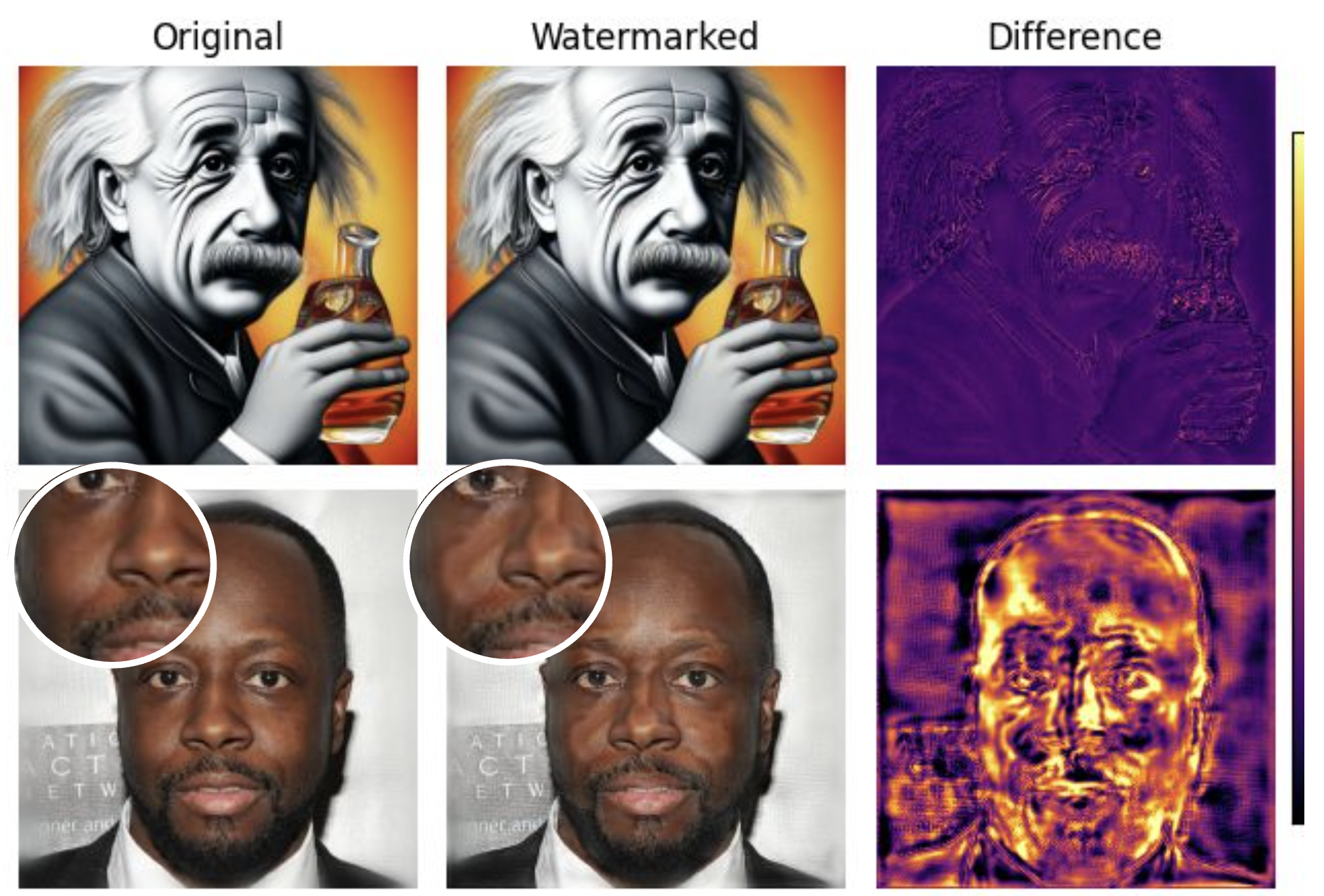}
  \caption{Images watermarked by \textit{StableSignature}--- \textit{non-semantic} (top), and \textit{StegaStamp}--- \textit{semantic} (bottom). The rightmost figures display the differences between the original and watermarked images that correspond to the changes encoding the watermarks. \textit{StableSignature}'s modifications are restricted to existing (high-frequency) edges such as wrinkles, hair, mustache, and intersections of multiple components. \textit{StegaStamp}'s watermark is distributed across the image. The magnified area shows how it manipulates the consistency (texture), injecting gradual (low-frequency) changes that manifest as wrinkles at this location.
  }
  \label{fig:example}
\end{figure}

We now turn our focus to the frequency distribution. There are two options for influencing frequencies to inject watermarks: Locally or globally. We claim the following: 

\noindent
\textbf{H2. \emph{Local changes to the pixel shift rates cannot reliably encode watermarks}}. The reason is that local modifications affect the rates at which pixels vary in select spatial regions. These share the above pitfall of strong dependency on the content, leaving only global spectral changes. The spectrum associates frequencies with two properties: \textit{magnitude} and \textit{phase}. Magnitudes represent the prevalence of shifts that occur at the specific rate across the image. Mathematically, any image is a superposition of periodic signals with varying frequencies occurring at different phases that determine their orientations, forming the image's structure based on how they intersect. Accordingly, and similar to the spatial structure, we deduce that: 
\textbf{H3. \emph{Phases are an \textit{unreliable carrier} due to their high sensitivity to the content}}. This is the result of the phases varying drastically between images as the primary factor affecting the structure (see Fig.~\ref{fig:spectrum}).  

By process of elimination, hypotheses \textbf{H1-H3} exclude all other candidates, leaving only one potential \textit{carrier}: the global spectral magnitudes. These are collective measurements with little dependency on specific pixels, making them resistant to reasonable manipulations that would preserve the majority of energy (amplitudes) distribution. They are also distributed similarly for any realistic content: lower frequencies contain the majority of energy as they correspond to gradual changes that capture the inherent consistency (continuity) of the worldly objects images depict. Higher frequencies (rapid transitions) occur at edges (intersections of objects), making them far less prevalent (see Fig.~\ref{fig:spectrum}). Thus, while amplitudes vary between images, they always follow a similar distribution, decaying as the frequencies grow. This stability other alternatives lack allows injecting watermarks via changes of similar magnitudes for any content, enabling them to be minimized s.t. they always integrate naturally.

\subsubsection{Spectral Watermarks}
\label{subsubsec:in_the_spec}
The above entails that robust watermarks reside in the global spectral amplitudes. Based on the different schemes' characteristics, we now present our main hypothesis, which serves as \textit{UnMarker}'s foundation: 

\noindent
\textbf{H4. \emph{\textit{Non-semantic} schemes primarily inject watermarks in the global spectral amplitudes associated with high frequencies, while \textit{semantic} watermarks mainly target low frequencies}}. Below, we analyze this claim:

\noindent
\underline{\textit{Non-Semantic:}} As these schemes cannot change the content, these are typically injected in high frequencies since sufficiently manipulating lower bands s.t. the watermark is detected requires substantial changes due to the vast energy concentration in this range. To alter the collective distribution of high frequencies, polarization at \textit{\textbf{existing}} edges can be controlled by adjusting the intensities of the various intersecting components to amplify or attenuate the transitions to their environments. This significantly impacts higher frequency amplitudes via small (invisible) modifications only due to the naturally low energy distribution in this range and by exploiting the characteristics of these regions (edges) that already exhibit high-frequency shifts to mask the watermark's footprint. Fig.~\ref{fig:example} (top) depicts this behavior for the \textit{non-semantic} \textit{StableSignature}~\cite{stablesig}. We verified that other \textit{non-semantic} schemes operate similarly. 

\noindent
\underline{\textit{Semantic:}} Since these schemes can influence the content, they will populate the consistency-defining low frequencies, making them more robust. As spectral amplitudes represent the \textit{collective} contribution to pixel variations occurring at the relevant rates from all spatial locations, these schemes construct their outputs s.t. the needed changes are spread among different regions, reducing visual cues. Unlike \textit{non-semantic} watermarks, they are not restricted to edges, affecting various objects' structures and textures. Since the changes required to alter low-frequency magnitudes are large, their influence may be visible. Fig.~\ref{fig:example} (bottom) portrays this for \textit{StegaStamp}~\cite{stegastamp}. \textit{TRW}'s~\cite{TreeRing} images may display different content from the original (non-watermarked), making illustrations uninformative (see~\S\ref{subsec:defensive}). Yet, our results in~\S\ref{subsec:unmarker_eval} prove its watermarks are in low frequencies.

\section{UnMarker}
\label{sec:unmarker}
Now that we have identified the \textit{carrier}, we present our universal attack. It follows from~\S\ref{subsubsec:candidates} (\textbf{H4}) that the \textit{carrier} must be a set $\mathcal{\boldsymbol{F}}_c=\{|F_b|:\; b\in [\![ B ]\!]\}$ of spectral amplitudes and the \textit{watermark property distribution}, denoted as ${\mathcal{\boldsymbol{F}}}_w$, is over these amplitudes. Let $\ell_d$ be a perceptual distance metric s.t. if $\ell_d(x,\; y) \leq t_{\ell_d}$ for some threshold $t_{\ell_d}$, then $x$ and $y$ are visually similar. We can now formalize watermark removal:
\begin{mytheorem} \textit{\underline{(WR):} Given watermarking algorithm $\mathcal{A}$ whose carrier frequency set is $\mathcal{\boldsymbol{F}}_c=\{|F_b|:\; b\in [\![ B ]\!]\}$ with watermark property distribution ${\mathcal{\boldsymbol{F}}}_w$, the watermark $w$ is removable from $x_w$ if there is an image $x_{nw}$ s.t. $\ell_d(x_w,\; x_{nw}) \leq t_{\ell_d} \land \mathcal{\boldsymbol{F}}_c(x_{nw}) \not\sim \mathcal{\boldsymbol{F}}_w$.} \label{t:wrt} \end{mytheorem}
Removing the watermark from $x_w$ amounts to finding a visually similar $x_{nw}$ for which $\mathcal{\boldsymbol{F}}_c$'s attributes deviate from the expected distribution $\mathcal{\boldsymbol{F}}_w$. Due to our black-box framework, the specific ${\mathcal{\boldsymbol{F}}}_c$ is unknown. 
 Instead, we seek $x_{nw}$ that retains visual similarity but maximizes some collective spectral difference $\ell_F$ from $x_w$:
\begin{equation}
\resizebox{0.55\hsize}{!}{
$\begin{aligned}
    x_{nw} = \underset{x^\prime}{argmax}&\; 
    [\ell_{F}(x_w,\; x^\prime)] \\[-5pt]
    &s.t.\; \ell_d(x_w,\; x^\prime) \leq t_{\ell_d}
\end{aligned}$
}
\label{eq:1}
\end{equation}

Maximizing the difference across all frequencies should lead to significant discrepancies in the values of ${\mathcal{\boldsymbol{F}}}_c$, forcing them to deviate from ${\mathcal{\boldsymbol{F}}}_w$. As \textit{semantic} and \textit{non-semantic} watermarks target different parts of the spectrum, we must craft a suitable attack strategy for each, identifying a proper loss function and optimization procedure.

\subsection{Attacking Non-Semantic Watermarks}
\label{subsec:high_freq}
We now instantiate \hyperref[eq:1]{\textit{eq.~(1)}} to defeat \textit{non-semantic} watermarks, while~\S\ref{subsec:semantic} presents a method against \textit{semantic} schemes. Following~\S\ref{subsubsec:candidates}, \textit{non-semantic} watermarks are in higher frequencies. Thus, we must disrupt this range without leaving visible cues. We present our high-frequency loss \textit{DFL} and the visual loss $\ell_d$, explaining how each step updates the input to arrive at a watermark-free output. 

\noindent
\textit{\textbf{Spectral Loss.}} We propose the \textit{Direct Fourier Loss (DFL)}:
\begin{equation}
\resizebox{0.75\hsize}{!}{
    $\forall\; x,y:\;\;\;\;\; DFL(x,\; y) = \|FT(x)-FT(y)\|_1$
}
\label{eq:lf0}
\end{equation}
where $FT$ is the ($2D$) Fourier transform. \textit{DFL} calculates the band-wise spectral differences between $x$ and $y$, outputting the sum of their magnitudes. Optimizing \textit{DFL} maximizes the amplitudes of the differences in the collective spectral representations between the sample being optimized and the watermarked image $x_w$, which leads to an output whose spectral magnitudes differ from $x_w$'s. Since the watermark is encoded in these amplitudes, it will be significantly disrupted, causing the detector to fail to recognize it. Yet, this is only guaranteed if the perturbation budget that controls the quality and similarity to the watermarked image is permissive enough for such significant distortions to occur. 
Luckily, this is true when disturbing high frequencies since the energy concentration in these bands, and accordingly, the watermark's footprint (as it should be invisible) is naturally small. Thus, watermarks become volatile, as they can be interfered with via minute perturbations.

\noindent
\textit{\textbf{Perceptual Loss.}} \textit{\underline{Learned Perceptual Losses (\textit{LPIPS}):}} They model human vision~\cite{lpips}. After experimenting with several candidates, we found 
\textit{DeepVGGLoss-LPIPS}~\cite{deeplossVGG} (\textit{DVL}) for large images ($\geq \text{\small {\mbox{$256 {\times} 256 $}}}$) and \textit{LPIPS-Alex}~\cite{lpips} (\textit{Alex}) otherwise optimal regardless of the scheme. High-frequency modifications may cause sporadic peaks to appear. \textit{Alex} fails to fathom the fragility of the larger images' higher resolution, often overlooking such singleton points despite being visible, while \textit{DVL} better captures this phenomenon. As these images are larger, numerous invisible violations throughout them sufficiently affect the spectrum. For smaller images, this restrictive loss prioritizes less destructive disruptions despite more powerful alternatives being applicable because of the reduced resolution. Due to the low dimensionality, such perturbations will not suffice, and increasing the budget results in visible cues as \textit{DVL} cannot distinguish between these nuances at this resolution. This problem is absent in low-frequency optimizations that are of a different nature (see~\S\ref{subsec:semantic}), where \textit{DVL} can be used with all sizes.

\noindent
\underline{\textit{Norm:}} To further ensure no such peaks appear (especially for low resolutions), we use the $\|\|_2$ norm to place a geometric bound on the distance from the watermarked image, allowing us to slightly increase the budget used for \textit{LPIPS}. Whenever \textit{LPIPS} risks introducing such peaks, $\|\|_2$ grows, forcing the procedure to explore different directions. Yet, $\|\|_2$ alone is insufficient since geometric constraints cannot capture visual similarity and will either lead to sub-optimal solutions if used with extremely small thresholds or impact quality otherwise. This is especially true since we operate in a \textit{black-box} framework without model feedback, potentially requiring larger perturbations, entailing that the perceptual loss should be accurate to preserve quality. 

\noindent
\textit{\textbf{Final High-Frequency Destruction Procedure.} (for non-semantic watermarks)} Using the above components, we implement \hyperref[eq:1]{\textit{eq.~(1)}}. We only need to identify the rules for updating the sample to define our search space. Since we seek high-frequency disruptions that affect the rates at which \textit{individual} pixels vary w.r.t. their neighbors, we should modify each pixel directly at each step. Thus, our algorithm accepts watermarked image $x_w$, \textit{LPIPS} loss $\ell_p$, \textit{LPIPS} threshold $t_{\ell_p}$, and $\|\|_2$ threshold $t_{\|\|_2}$, and finds modifier $\delta$ that maximizes \textit{DFL} as follows:
\begin{equation}
\resizebox{\hsize}{0.45cm}{
$x_{nw} = x_w + \underset{\delta}{argmin}\;      \left[
        \begin{array}{cc}
           c_{\ell_p} \cdot ReLU_{\ell_p}(x_w+\delta,\; x_w) 
           +c_{\|\|_2} \cdot ReLU_{\|\|_2}(x_w+\delta,\; x_w) \\[5pt] -
           DFL(x_w+\delta,\; x_w)
        \end{array}
        \right]$
}
\label{eq:5}
\end{equation}

$ReLU_{\ell}(x,\; y)$ denotes $max\{\ell(x,\; y)-t_{\ell}, 0\}$ for any loss $\ell$. Effectively, we instantiate $\ell_d$ from \hyperref[eq:1]{\textit{eq.~(1)}} with two components: \textit{LPIPS} loss $\ell_p$ and norm $\|\|_2$, each with its threshold, while \textit{DFL} is the spectral loss. The vector $\delta$ captures the changes to $x_w$ to generate the non-watermarked $x_{nw}\!\!\!=\!\!x_w\!\!+\!\!\delta$. We minimize -\textit{DFL}, maximizing \textit{DFL}. The constants $c_{\ell_p}$ and $c_{\|\|_2}$ balance the spectral and perceptual losses. We enforce that each perceptual loss ($\ell_d$) component ($\ell_p$ and $\|\|_2$) remains below its threshold ($t_{\ell_p}\!$ or $t_{\|\|_2}$) through the term $ReLU_{\ell_d}(x_w\!\!+\delta,\; x_w)$: When $\ell_d$ exceeds $t_{\ell_d}$, it evaluates to $\ell_d(x_w\!+\delta,\; x_w)-t_{\ell_d}$, causing $\ell_d$'s gradients to produce a sample closer to $x_w$ in the next iteration. Otherwise, it returns $0$, discarding $\ell_d$'s gradients and maximizing \textit{DFL} unconditionally. $c_{\|\|_2}$ is a fixed \textit{large} coefficient that strongly prevents the abovementioned peaks. $c_{\ell_p}$ is determined via a binary search: We start with a large $c_{\ell_p}$, assigning a larger weight to $\ell_p$'s gradients and run the optimization for several steps. If a solution is found, $c_{\ell_p}$ is decreased, increasing \textit{DFL}'s dominance at the expense of larger visual deviations. Otherwise, $c_{\ell_p}$ is amplified. See  Appendix~\ref{app:hyper} for details.

\subsection{Attacking Semantic Watermarks} 
\label{subsec:semantic}
\textit{Semantic} watermarks are in low frequencies, making them difficult to address. In addition to their large magnitudes (see~\S\ref{subsubsec:in_the_spec}), low frequencies correspond to many neighboring pixels varying gradually, indicating these watermarks are rooted in the image's texture. Thus, removing them requires \textit{systemic} and structural changes. Direct optimizations similar to~\S\ref{subsec:high_freq} fail since they operate on the image and, at each step, individual pixels are updated in different directions. This targets higher frequencies, introducing significant changes yet \textit{\textbf{only}} to a few pixels, failing to progress due to the perceptual constraints and incurring minimal changes at low frequencies. We could limit the procedure to optimize \textit{DFL} over low-frequencies only. Yet, as significant changes are required in this range, without a methodical scheme, we will converge to a sub-optimal solution wherein a certain region receives substantial modifications, leading the perceptual loss to prevent further updates. This is because the process fails to account for inter-pixel links, and \textit{DFL} does not spread the changes over the image.

\begin{figure*}[!t]
\begin{minipage}{0.45\textwidth}
\centering
\subfloat[Basic: To calculate the output at $(i,\; j)$, the patch $PE^{i, j}_{\text{\tiny \mbox{$M \!{\times} \! N$}}}$ (where {\small \mbox{$M \! {\times} \! N$}} is the kernel size) around $(i,\; j)$ is element-wise multiplied by the \textit{spectral kernel} $\mathcal{K}$, summing over this product.]{\includegraphics[width=\linewidth,height=2.3cm]{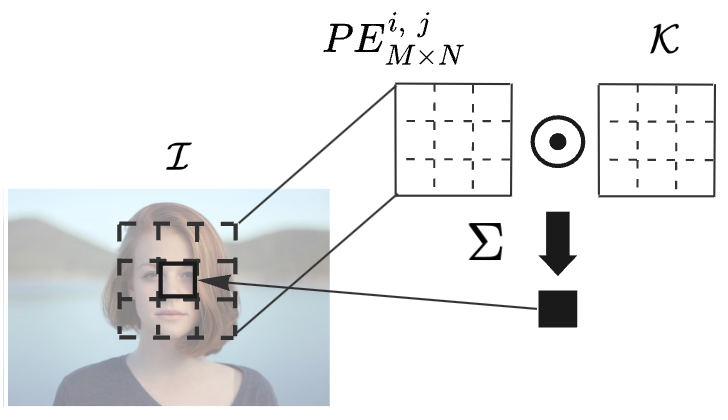}}%
\end{minipage}%
\hfill
\begin{minipage}{0.45\textwidth}
\centering
\subfloat[Guided: The \textit{color kernel} {\small \mbox{$\mathcal{C}^{\mathcal{I},\sigma\!_{_c}}_{i,j}$}} is first computed, assigning a modulation factor to each neighbor in the vicinity $PE^{i,j}_\text{{\tiny \mbox{$M \! {\times} \! N$}}}$ of $(i,\; j)$. These are element-wise multiplied by the \textit{spectral kernel} $\mathcal{K}$ to limit the contribution from neighbors whose values differ significantly from $(i,\; j)$'s. This yields the \textit{effective kernel} $\mathcal{V}^{\mathcal{I},\mathcal{K},\sigma\!_{_c}}_{i,j}$ used to calculate the output at $(i,\; j)$ similar to basic filters.]{\includegraphics[width=\linewidth,height=2.4cm]{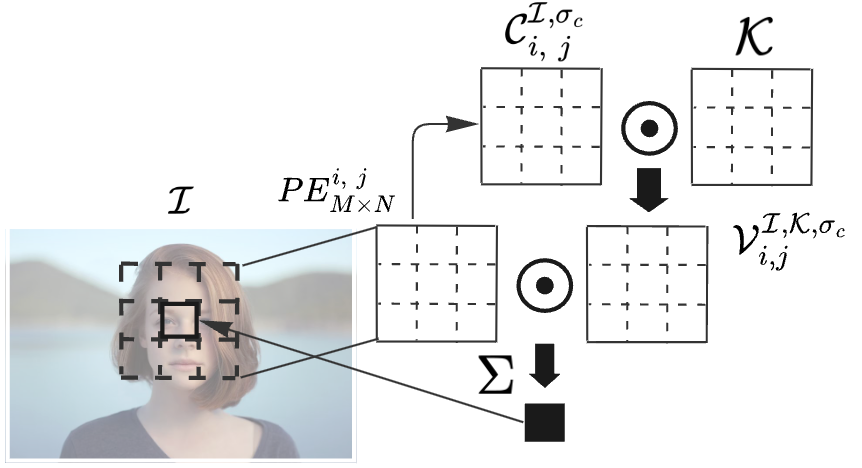}}%
\end{minipage}%
\caption{Operation of known filters.}
\label{fig:filters}
\end{figure*}

We develop a visually-aware attack inspired by image filtering. While filters apply systemic spectral disruptions, they fail to remove watermarks due to their content-agnostic nature, limiting the distortions they may introduce without leaving crucial visual cues. Our adversarial alternative retains quality by learning the filters' weights to solve an optimization problem that maximizes low-frequency disruptions, resorting to a perceptual loss to preserve critical information. First, we motivate our design through guiding principles. We then provide technical details and present our optimizable filters. Finally, we introduce our attack.

\subsubsection{Motivation}
\noindent
\textit{\underline{Basic Filtering}:} An image is convolved with a filter ($\text{\small\mbox{$M {\times} N$}}$ \textit{spectral kernel}), moved s.t. its center is aligned with a single pixel $(i,\; j)$ at each step. The output at $(i,\; j)$ is calculated by multiplying the kernel by the pixel's vicinity and summing over the product--- see Fig.~\ref{fig:filters}(a). The kernel's values are non-negative and sum to $1$, making this output an averaged sum of the pixel and its neighbors. This performs noise reduction, decaying unwanted peaks. The kernel's elements grow toward the center, assigning a larger weight to the original pixel to preserve the content. 

Convolution in space translates to multiplication in the frequency domain: Given the spectra of the image and filter, it produces an image with a spectrum equivalent to their product. Thus, filters are good candidates for spectral manipulations. Yet, the outputs at any pixel are calculated using the same kernel. Hence, their potential to remove watermarks is limited~\cite{ptw,provable} due to their content-agnostic nature, making them only applicable with mild parameters as stronger blurring impacts quality. 

\noindent
\textit{\underline{Guided Filtering:}} These advanced filters tackle the information loss under stronger distortions, allowing more effective noise reduction while preserving crucial data. They are edge-preserving, minimizing the disruptions at these critical boundary points between different objects. Instead of using a uniform kernel, they involve an additional per-pixel \textit{color kernel} that weakens the filter at edge points based on geometric principles: If some values in the pixel's vicinity drastically differ from it, then it lies on an edge, as these neighbors belong to other components and the \textit{spectral kernel}'s strength here must be diminished to preserve defining information. The \textit{color kernel} computes a modulation factor for every neighbor in the filter's vicinity around $(i,\; j)$ based on its value's difference from $(i,\; j)$'s. The \textit{effective kernel} for calculating $(i,\; j)$'s output is the Hadamard product of the two kernels--- see Fig.~\ref{fig:filters}(b). Guided filters fail to remove watermarks despite their stronger blurring (see~\S\ref{subsubsec:utility}) since they merely follow geometric heuristics that fail to fully capture human vision or target specific image characteristics, limiting the allowed distortion. In practice, less \textit{visually-critical} areas can be changed drastically, regardless.

\noindent
\textit{\underline{Adversarial Filtering} (Our Attack):} Based on the above, structure-aware filters can better maximize the spectral distortion with minimal impact on crucial regions. Similar to guided filters, we design a process for learning a set of per-pixel kernels. Yet, instead of relying on geometric heuristics, our method learns the kernels' parameters s.t. they retain \textbf{\textit{perceptual similarity}} to the original image from a human's perspective, allowing for stronger filtering. 

\noindent
\subsubsection{Formalizing Filters}
Given image $\mathcal{I} \!\! \in \!\! \mathbb{R}^{\text{\tiny{ \mbox{$Q \! {\times} \! R$}}}}$, $PE^{i, j}_{\text{\tiny \mbox{$M \! {\times} \! N$}}}: \mathbb{R}^{\text{\tiny{ \mbox{$Q \! {\times} \! R$}}}}$ $ \!\! \longrightarrow \!\! \mathbb{R}^{\text{\tiny{ \mbox{$M \! {\times} \! N$}}}}$ denotes the \textit{local patch extractor} that outputs the \mbox{\small\text{$M {\times} N$}} vicinity around its $(i,\; j)^{th}$ pixel s.t. this pixel is at the center (i.e., $PE^{i, j}_{\text{\tiny \mbox{$M \! {\times} \! N$}}}(\mathcal{I})(\text{\small \mbox{$\floor{\frac{M}{2}}$}},\; \text{\small \mbox{$\floor{\frac{N}{2}}$}}) = \mathcal{I}(i,\; j)$)--- see Fig.~\ref{fig:filters}(a). The \textit{global patch extractor} $PE_{\text{\tiny \mbox{$M \! {\times} \! N$}}}: \mathbb{R}^{\text{\tiny{ \mbox{$Q {\times} R$}}}} \!\! \longrightarrow \!\! \mathbb{R}^{\text{\tiny{ \mbox{$Q \! {\times} \! R {\times} M \! {\times} \! N$}}}}$ outputs a matrix that stores in cell $(i,\; j)$ the \textit{local extractor} $PE^{i,j}_{\text{\tiny \mbox{$M \! {\times} \! N$}}}$. That is, $PE_{\text{\tiny \mbox{$M \! {\times} \! N$}}}(\mathcal{I})$'s $(i,\; j)^{th}$ cell contains the \mbox{\small$M {\times} N$} patch around $(i,\; j)$ in $\mathcal{I}$. We formalize filters to present our solution.

\noindent
\textit{\underline{Basic Filtering}:}
Basic filters simply apply the known convolution with a fixed \textit{spectral kernel} $\mathcal{K} \in \mathbb{R}^{\text{\tiny{ \mbox{$M \! {\times} \! N$}}}}$. Denoting element-wise multiplication as $\odot$, these filters can be written as $\mathcal{F}_\mathcal{K}: \mathbb{R}^{\text{\tiny{ \mbox{$Q \! {\times} \! R$}}}} \! \longrightarrow \! \mathbb{R}^{\text{\tiny{ \mbox{$Q \! {\times} \! R$}}}}$ s.t. $\mathcal{F}_\mathcal{K}
(I)(i,\; j) = \sum_{m,n} PE^{i,j}_{\text{\tiny \mbox{$M \! {\times} \! N$}}}(I) \odot \mathcal{K}$ (see Fig.~\ref{fig:filters}(a)).

\noindent
\textit{\underline{Guided Filtering:}} These employ additional per-pixel \textit{color kernel}s to modulate distortions at critical points (Fig.~\ref{fig:filters}(b)). Given kernel size \mbox{\small$M {\times} N$}, the \textit{color kernel}  for each $(i,\; j)$ is the matrix whose $(m,\; n)$ cell is {\small \mbox{$\mathcal{C}^{\mathcal{I},\sigma\!_{_c}}_{i,j}(m,\; n)$}} = $e^{\mbox{\small${-\frac{\|\mathcal{I}(i+m-\floor{\frac{M}{2}},\; j+n-\floor{\frac{N}{2}})-\mathcal{I}(i,\; j)\|_1}{2 \sigma\!_{_c}^2}}$}}$. $\sigma\!_{_c}$ is a permissiveness controlling the modulation intensity. The {\small \mbox{$\|\|_1$}} norm accounts for each pixel potentially having multiple color channels. For instance, \textit{RGB} images associate each pixel with three values. The \textit{effective kernel} for computing the output at $(i\; j)$ is then $\mathcal{V}^{\mathcal{I},\mathcal{K},\sigma\!_{_c}}_{i,j} = \text{\small \mbox{$\frac{\mathcal{C}^{\mathcal{I},\sigma\!_{_c}}_{i,j} \odot \mathcal{K}}{\underset{m,n}{\text{\large \mbox{$\sum$}}}\mathcal{C}^{\mathcal{I},\sigma\!_{_c}}_{i,j} \odot \mathcal{K}}$}}$. At the center, {\small \mbox{$\mathcal{C}^{\mathcal{I},\sigma\!_{_c}}_{i,j}(\floor{\frac{M}{2}},\;\floor{\frac{N}{2}})=1$}}, indicating the contribution of the original $\mathcal{I}(i,\; j)$ is not decayed. For neighbors, large distances decrease their coefficients and limit their contributions. If we define $\mathcal{V}^{I,\mathcal{K},\sigma\!_{_c}}$ as the matrix whose $(i,\; j)^{th}$ cell contains $\mathcal{V}^{\mathcal{I},\mathcal{K},\sigma\!_{_c}}_{i,j}$, the guided filter is given as {\small \mbox{$GF_{\mathcal{K}, \sigma_c}(\mathcal{I}) = \underset{m,n}{\text{\large \mbox{$\sum$}}} PE_{\text{\tiny \mbox{$M \! {\times} \! N$}}}(\mathcal{I}) \odot \mathcal{V}^{\mathcal{I},\mathcal{K},\sigma\!_{_c}}$}}.

\noindent
\underline{\textit{Optimizable Filters} (Ours):} Guided filters' drawback is that they employ the same \textit{spectral kernel} $\mathcal{K}$ for all pixels, changing only the \textit{color kernel} computed via the same formula at all locations. Instead, we allow each pixel $(i,\; j)$ to have a unique \textit{spectral kernel} ${\mathcal{K}\!\!\!_{\;\;i,j}}$, with weights that can differ significantly between locations based on feedback from a perceptual loss that determines their visual importance and allows maximizing the destruction at less critical regions. Our filters consist of a set $\mathcal{K}^*$ of local ${\mathcal{K}\!\!\!_{\;\;i,j}}$'s. We include \textit{color kernel}s to avoid exploring unwanted directions that violate vital geometric constraints. While these may be limiting, ignoring them causes unnatural distortions if the differences between neighboring pixels are extremely large, making them essential for remedying potential perceptual loss imperfections. When the distances are reasonably large but satisfy the perceptual loss with high confidence, it will assign larger weights that counteract the \textit{color kernel}. Only when these distances are extremely large do they become prohibitive, which is ideal. Our optimizable filter is:
\begin{equation}
\resizebox{0.75\hsize}{!}{
$\forall \mathcal{I}:\;  OF_{\mathcal{K}^*,\mathcal{R}, \sigma\!_{_c}}(\mathcal{I}) = \underset{m,n}{\text{\large \mbox{$\sum$}}} PE_{\text{\tiny \mbox{$M {\times} N$}}}(\mathcal{I}) \odot \mathcal{V}^{\mathcal{R},\mathcal{K}^*,\sigma\!_{_c}}$
}
\label{eq:optimizable_filter}
\end{equation}
$\mathcal{R}$ is a reference image for obtaining the \textit{color kernel}s, set to $\mathcal{R}\equiv x_w$. We now show how to learn $\mathcal{K}^*$'s weights.

\subsubsection{Adversarial Filtering}
\label{subsec:optimizable_filts}
Let \mbox{\small$\{t \in [\![ T ]\!]|OF_{\mathcal{K}^*\!\!\!_{_t}, x_w, {\sigma\!_{c_{_t}}}}\}$} as a series of $T$ optimizable filters (with \textit{color kernel}s computed using $x_w$), each with their permissiveness and size. We define the composition $\mbox{\small$\overset{T}{\underset{\boldsymbol{OF}}{\prod}}$} \! \equiv \!\!\! {\text{\small \mbox{$\mathrel{\raisebox{3pt}{$OF_{\mathcal{K}^*\!\!\!_{_1}, x_w, {\sigma\!_{c_{_1}}}} \! \circ \! OF_{\mathcal{K}^*\!\!\!_{_2}, x_w, {\sigma\!_{c_{_2}}}} \!\!\! \circ \cdot\cdot\cdot \circ OF_{\mathcal{K}^*\!\!\!_{_T}, x_w, {\sigma\!_{c_{_T}}}}$}}$}}}$. Let $\ell_d$ be a perceptual loss, $t_{_{\ell_d}}$ a similarity threshold, and $\ell_F$ a low-frequency loss. We instantiate \hyperref[eq:1]{\textit{eq.~(1)}} to maximize $\ell_F$ as:
\begin{equation}
\resizebox{0.7\hsize}{!}{
$
\begin{aligned}
    x_{nw} = \underset{\{\theta_{\mathcal{K}^*\!\!\!_{_t}}\}, \delta}{argmax}&\; 
    [\ell_{F}(x_w,\; \mbox{\small$\overset{T}{\underset{\boldsymbol{OF}}{\prod}}$}(x_w+\delta))] \\[-8pt]
    &s.t.\; \ell_d(x_w,\; \mbox{\small$\overset{T}{\underset{\boldsymbol{OF}}{\prod}}$}(x_w+\delta))) \leq t_{\ell_d}
\end{aligned}
$
}
\label{eq:struct}
\end{equation}
where $\theta_{\mathcal{K}^*\!\!\!_{_t}}$ are the kernels' weights and $\delta$ is a modifier that directly optimizes $x_w$, similar to \hyperref[eq:5]{\textit{eq.~(2)}}. We maximize the spectral loss by sequentially propagating the input through a network of filters. Recall that direct modifications are not effective for disrupting low frequencies in the absence of filters due to their inclination to cause high-frequency disturbances rather than systemic shifts in the values of many neighboring pixels in a gradual manner--- see~\S\ref{subsec:semantic}. Thanks to their convolutional nature, this no longer holds with our filters, making their outputs combinations of neighboring pixels. We still include $\delta$ that directly manipulates the input despite its inefficacy in the absence of filters. This is because the originally high-frequency disturbances caused by such modifications will now be distributed over neighbors of the receiving pixels due to our filters, translating them into useful low-frequency noise, causing further disruptions beyond the filters' independent ability. We employ multiple ($T$) filters stacked after one another. The reason is that chaining several filters with various kernel sizes (geometries) targets different directions and slopes where the watermark could have been injected since watermarking is content-dependent and not all images can be changed similarly.

\subsubsection{Spectral Loss}
\label{subsubsec:low_freq_loss}
The former \textit{DFL} is unsuitable for low frequencies (see~\S\ref{subsec:semantic}) as it fails to force the changes to spread throughout the image. The collective energy at higher frequencies is much smaller in comparison, significantly reducing the total number of pixels that need to change to erase watermarks in these bands, which is achievable without violating the perceptual constraints. In contrast, sufficiently manipulating lower frequencies can rapidly change the structure, hindering convergence to an optimal solution if the distortion is not carefully distributed. This loss should also ensure our kernels perform the filtering operation, which \textit{DFL} fails to enforce. Better alternatives are:

\noindent
\textit{\underline{Filter Regularization Loss} (\textit{FRL}):} 
It enforces that our kernels indeed operate as (smoothing) filters and is given as:
\begin{equation*}
\resizebox{0.65\hsize}{0.65cm}{ 
$\begin{aligned}
    &\forall x \in \mathbb{R}^{\text{\tiny{ \mbox{$Q \! {\times} \! R$}}}}:\; \textit{\text{FRL}}(x) = -\sum_{t \in [\![ T ]\!]} \textit{\text{FRL}}_{_{\mathlarger{\boldsymbol{t}}}}(x) \\[-5pt]
    &\;\;\;\;\;\;\;\;\;\;\;s.t.\;\forall t:\;\; \textit{\text{FRL}}_{_{\mathlarger{\boldsymbol{t}}}}(x) = \|x-{{\mathlarger{\mathlarger{\mathlarger{\mu}}}}\!_{_{_{\frac{1}{2}}}}\!\!\!\!\!\!\!}^{^{^{\text{\tiny \mbox{$M_t \! {\times} \! N_t$}}}}}\!\!\!\!\!\!(x)\|_2
\end{aligned}$
}
\end{equation*}
{\small \mbox{$M_t {\times} N_t$}} is the size of our $t^{th}$ kernel ${\mathcal{K}^*\!\!\!_{_t}}$ and ${{\mathlarger{\mathlarger{\mathlarger{\mu}}}}\!_{_{_{\frac{1}{2}}}}\!\!\!\!\!\!\!}^{^{^{\text{\tiny \mbox{$M_t \! {\times} \! N_t$}}}}}\!\!\!\!\!\!$ computes the median of this vicinity around each pixel. Maximizing \textit{FRL} minimizes $\textit{\text{FRL}}_{_{\mathlarger{\boldsymbol{t}}}}$ for every filter $t$, which is the sum of all pixels' distances from the median in the filter's environment around each pixel, bringing pixels closer to these local medians. This eliminates variations to the extent possible under the perceptual loss, leading to smoother outputs. As the watermark is distributed over the image, such patches will exhibit gradual ``watermarking'' variations that will now be lost. A major benefit of \textit{FRL} is output consistency.

\noindent
\textit{\underline{Mean Pool Loss} (\textit{MPL}):}
\textit{FRL} is insufficient, ensuring only proximity of the pixels to the local medians but imposing no constraints on these medians. The \textit{means} of the various patches in the \textit{watermarked} image are highly influenced by the watermark that causes many pixels to change their values. Due to the perceptual loss that strives to retain information, the procedure may converge to a sub-optimal solution, skewing these medians toward the original \textit{means} and only slightly affecting pixel intensities, retaining many shifts associated with the watermark. Thus, we require \textit{MPL}:
\begin{equation*}
\resizebox{0.7\hsize}{0.3cm}{
$\forall x,y \in \mathbb{R}^{\text{\tiny{ \mbox{$Q \! {\times} \! R$}}}}:\;\;
    \textit{\text{MPL}}^{^{\text{\tiny \mbox{$L \! {\times} \! P$}}}}(x,\; y) = \|\overline{x}^{^{\text{\tiny \mbox{$L \! {\times} \! P$}}}} \!\!\! - \overline{y}^{^{\text{\tiny \mbox{$L \! {\times} \! P$}}}}\|_1$
}
\end{equation*}
$\overline{x}^{^{\text{\tiny \mbox{$L \! {\times} \! P$}}}}\!\!$ denotes mean pooling with an {\small \mbox{$L {\times} P$}} window. 
\textit{MPL} takes the means over each {\small \mbox{$L {\times} P$}} patch in both inputs and calculates the {\small \mbox{$\ell_1$}} difference. As most pixels do not encode watermark information, maximizing the means' differences (w.r.t. the watermarked image) while demanding proximity to the medians (using \textit{FRL}) primarily affects pixels located far (those that encode the watermark) from all others, bringing them closer to the rest and eliminating watermarking variations. A proper low-frequency loss must spread the changes over multiple regions (see~\S\ref{subsubsec:low_freq_loss}). \textit{MPL} does so by dividing the image into patches and demanding that each receive a portion of the change, implying the number of patches must be large. Yet, each must contain enough pixels to manipulate \textit{low} frequencies. We choose ${\text{\small \mbox{\textit{MPL}}}}^{^{\text{\tiny \mbox{$5 \! {\times} \! 5$}}}}\!\!$ as a compromise, balancing these requirements. Calculating \textit{MPL} for all filter sizes similar to \textit{FRL} is less desirable as when the sizes shift away from this sweet spot, one of the conditions is violated. Note that, without \textit{FRL}, \textit{MPL} is insufficient as it can lead to all pixels in each patch receiving a similar modification $\epsilon$, simply changing its brightness. This satisfies the perceptual constraints but leaves the spectrum unaffected as the pixel variation rates remain identical.

\subsubsection{Final Low-Frequency Destruction Procedure} \textit{(for semantic watermarks)}
\noindent
We replace \hyperref[eq:struct]{\textit{eq.~(5)}}'s abstractions with the above components and $\ell_d=$ \textit{{\small \mbox{DVL}}} as the perceptual loss (see~\S\ref{subsec:high_freq}). We also use $\|\|_2$ regularization to avoid drastic updates and encourage low-frequency, gradual changes. Yet, it is multiplied by a small coefficient to only regulate the process without further influencing the outputs, leaving this to \textit{{\small \mbox{DVL}}}. Defining {\small \mbox{$x_w^{^{OF_{_{\text{\tiny \mbox{$T$}}}},\delta}} \equiv \mbox{\small$\overset{T}{\underset{\boldsymbol{OF}}{\prod}}$}(x_w+\delta)$}}, we have: 
\begin{equation}
\resizebox{\hsize}{0.5cm}{ 
    $x_{nw} = \; \underset{\{\theta_{\mathcal{K}^*\!\!\!_{_t}}\}, \delta}{argmin}\; 
    \left[
        \begin{array}{cc}
    c_{{{\text{\tiny \mbox{${DVL}$}}}}} \cdot ReLU_{{{\text{\tiny \mbox{${DVL}$}}}}}(x_w,\; x_w^{^{OF_{_{\text{\tiny \mbox{$T$}}}},\delta}}) + \lambda_{{\|\|_2}}\cdot \|x_w^{^{OF_{_{\text{\tiny \mbox{$T$}}}},\delta}}-x_w\|_2\\[5pt]
    -MPL^{^{\text{\tiny \mbox{$5 {\times} 5$}}}}(x_w,\; x_w^{^{OF_{_{\text{\tiny \mbox{$T$}}}},\delta}})-\lambda_{{\text{\tiny \mbox{$FRL$}}}}\cdot FRL(x_w^{^{OF_{_{\text{\tiny \mbox{$T$}}}},\delta}})
    \end{array}
    \right]$
}
\label{eq:struct_fin}
\end{equation}
We maximize the loss $\ell_{F}\! = \!\!{\text{\textit{MPL}}}^{^{\text{\tiny \mbox{$5 {\times} 5$}}}} \!\!\!\! + \lambda_{{\text{\tiny \mbox{$FRL$}}}}\cdot {\text{\textit{FRL}}}$ by minimizing its negation similar to~\S\ref{subsec:high_freq}. $ReLU_{{{\text{\tiny \mbox{${DVL}$}}}}}$ ensures similarity, keeping the \textit{DVL} distance from $x_w$ below $t_{{{\text{\tiny \mbox{${DVL}$}}}}}$. The resulting changes are gradual due to our filters, whose weights are learned to achieve this behavior thanks to our spectral loss, yielding optimal {\small \mbox{$\{\widehat{\theta}_{\mathcal{K}^*\!\!\!_{_t}}\}, \widehat{\delta}$}} s.t. the final solution is $x_{nw}\!\!=\!\!\text{\mbox{$x_w^{^{\widehat{OF}_{_{\text{\tiny \mbox{$T$}}}},\widehat{\delta}}}$}}\!\!\!\!$, where {\tiny \mbox{$\widehat{OF}_{_{\text{\tiny \mbox{$T$}}}}$}} denotes the filters applied using {\small \mbox{$\{\widehat{\theta}_{\mathcal{K}^*\!\!\!_{_t}}\}$}}. As each kernel's values should be non-negative and sum to $1$, we apply \textit{{\small \mbox{Softmax}}} to the learned weights to obtain effective values in this range. $\lambda_{{\text{\tiny \mbox{$FRL$}}}}$ balances \textit{FRL} and \textit{MPL} and is calibrated s.t. both are optimized, while $\lambda_{{\|\|_2}}$ limits the impact of the $\|\|_2$ regularization. $c_{{\text{\tiny \mbox{$DVL$}}}}$ balances the spectral and visual losses and is determined via a binary search, as in~\S\ref{subsec:high_freq}. We found a single filter network architecture universally effective. Details are in Appendix~\ref{app:hyper}.

\noindent
\textit{\underline{Discussion:}} Without our filters, the modifications at each step are less systemic, allowing different pixels to receive significant updates in opposing directions. Thus, we will converge to a sub-optimal solution since massive changes that a \textit{subset} of pixels receive will substantially increase the visual loss, preventing further progress. While our low-frequency loss discourages this behavior and encourages uniform updates, it remains lacking, constantly countered by the visual loss that strives to retain similarity, limiting the number of pixels allowed to change their values, deeming such structural modifications less desirable compared to high-frequency changes that affect discrete pixels thereby remaining potentially unnoticeable. Our filters force prioritizing collective updates, drastically improving performance. Results establishing these benefits are in Appendix~\ref{app:filt}. 

While effective against \textit{semantic} watermarks, this attack is unsuitable for \textit{non-semantic} ones (see~\S\ref{subsec:unmarker_eval}); The low-frequency loss and filters encourage low-magnitude, yet collective, changes and prevent significant updates to specific pixels (as opposed to~\S\ref{subsec:high_freq}). Yet, \textit{non-semantic} watermarks are injected in high frequencies and thus rely on large variations between neighboring pixels. As this process can only slightly affect such peaks, the watermark is preserved.

%% file: Sections/Experiments.tex
\section{Experiments}
\label{sec:exp}

\subsection{Setup}
\label{subsec:setup}

\noindent
\textit{\textbf{Schemes.}} We consider the \textit{SOTA} watermarking schemes in Table~\ref{tab:wat_schemes}, covering all categories from~\S\ref{sec:Background}, selected for appearing in top venues, with several remaining unbroken by a \textit{query-free black-box} attacker. Among the \textit{generator-specific} schemes, \textit{Yu1}~\cite{yu1}, \textit{Yu2}~\cite{yu2}, and \textit{PTW}~\cite{ptw} watermark \textit{Generative Adversarial Networks (GANS)}~\cite{GAN}, while \textit{StableSignature}~\cite{stablesig} and \textit{TRW}~\cite{TreeRing} operate on \textit{LDMs}~\cite{latentdiff}. We are not aware of \textit{semantic} schemes from top venues other than those we consider, which are the focus of previous work~\cite{provable,saberi2023robustness,WAVES}. As most recent systems are \textit{generator-specific}, we include five such schemes and two \textit{general-purpose} ones. Compared to existing works (see~\S\ref{subsec:attacks_backround}), we span a wider variety of recent systems.

\begin{table}[h!]
\caption{Studied \textit{SOTA} watermarking schemes.}
\label{tab:wat_schemes}
\renewcommand{\arraystretch}{1.}
\fontsize{21pt}{21pt}\selectfont
\resizebox{\columnwidth}{0.7cm}{
\begin{tabular}{l"c|c}
\thickhline
\addlinespace[10pt]
& \textit{\textbf{General-Purpose}} & \textit{\textbf{Generator-Specific}}\\
\midrule
\addlinespace[10pt]
\textit{\textbf{Semantic}} & \textit{StegaStamp~\cite{stegastamp}} & \textit{TRW~\cite{TreeRing}} \\ [10pt]
\textit{\textbf{Non-Semantic}} & \textit{HiDDeN~\cite{HiDDeN}} & \textit{Yu1~\cite{yu1}}, \textit{Yu2~\cite{yu2}}, \textit{PTW~\cite{ptw}}, \textit{StableSignature~\cite{stablesig}} \\
\thickhline
\end{tabular}
}
\end{table}

\noindent
\textit{\textbf{Data.}} Most \textit{generator-specific} schemes require no input data. They were trained on various datasets: \textit{TRW} and \textit{StableSignature} on \textit{LAION-5B}~\cite{Laion}, \textit{Yu1} and \textit{Yu2} on \textit{CelebA}~\cite{progan}, and \textit{PTW} on \textit{FFHQ}~\cite{stylegan}. As \textit{TRW} and \textit{StableSignature} watermark text-to-image \textit{LDM}s (\textit{StableDiffusion}~\cite{Clipdrop}), we use \textit{StableDiffusion}'s prompts. For \textit{HiDDeN}, we take samples from \textit{COCO}~\cite{coco} on which it was trained. \textit{StegaStamp} watermarks {\small \mbox{$400 {\times} 400$}} images. Thus, we downscale \textit{CelebA-HQ-1024} samples\footnote{\textit{FFHQ-1024} samples yield similar results (excluded for brevity).}. We evaluate each system on $100$ random images, similar to previous work~\cite{WeVade}.

\noindent
\textit{\textbf{Metrics.}} Following~\S\ref{sec:Threat}, we are concerned with two metrics:

\textit{1) Detection:} \emph{\underline{Bit Accuracy.}} For binary watermarks, detection occurs if the ratio of bits matching the watermark in the sequence extracted from the image is above some threshold $t_w$. As in previous work~\cite{provable,lukas2023leveraging}, we select $t_w$ that can reject the null hypothesis $H_0$ that $k$ or more matching bits
were randomly extracted with $p<0.01$. This has probability $P(X \geq k| H_0) = \sum_{i=k}^n \binom{n}{k} \frac{1}{2^n}$, for watermark length $n$. Thus, $t_w=\frac{k}{n}$ for which this probability is below $0.01$ is selected. We denote the portion of successfully extracted bits as the \textit{bit accuracy}, which we compare against $t_w$. Watermark lengths are in Appendix~\ref{app:wat_len}. 

\noindent
\emph{\underline{Inverse Distance.}} \textit{TRW}'s watermarks are unbounded, and detection occurs if the $\ell_1$ distance (effectively, the \textit{Mean Absolute Error--- MAE}) of the extracted sequence from the watermark is below $t_w$ determined s.t. the ratio of falsely detected non-watermarked images (\textit{FPR}) is low. We chose a higher \textit{FPR} ($0.02$) than the original paper~\cite{TreeRing} ($0.01$), making the attacker's task harder, obtaining $\textit{\text{MAE}}=71$. For consistency, we use this distance's \textit{inverse}: for the watermark to be detected, the \textit{MAE}'s \textit{inverse} must be $\geq t_w=\frac{1}{71}$.

\noindent
\emph{\underline{Security Advantage.}} This \textit{binary} indicator is the scheme's ability to detect over $50\%$ of \textit{watermarked} images. 
Most genuine content is not controversial, and detectors will not be invoked to determine its authenticity. They are only necessary when suspicious content surfaces. Absent additional knowledge, we assume a natural-looking controversial image has equal probabilities of being real or fake. Here, the best \textit{trivial} detector that achieves the highest average accuracy on suspicious images would be a random classifier. An attacker will have a $50\%$ chance of fooling it with a fake image. For a scheme to have a \textit{Security Advantage}, it must detect \textit{watermarked} images at a higher rate. Otherwise, the resources invested in it will be in vain as it offers no additional protection against attackers compared to random systems. While such a scheme will perform better on non-watermarked images as it is trained to detect the existence of a watermark, this is irrelevant from a security aspect where only identifying the fake (watermarked) content matters.

\textit{2) Quality}: We use two common ~\cite{provable,WAVES,lukas2023leveraging,hu2024transfer,ptw} human vision-inspired metrics: \textit{LPIPS}~\cite{lpips} and \textit{Fr\`echet Inception Distance (FID)}~\cite{fid}. For fairness, we avoid \textit{LPIPS} methods \textit{UnMarker} explicitly optimizes (see~\S\ref{subsec:high_freq}), resorting to \textit{LPIPS-VGG}~\cite{lpips}. \textit{FID} compares the attack samples' distribution to that of watermarked images, measuring \textit{dataset-wise} similarity. We exclude geometric metrics (\textit{PSNR} or norms) as we find them misleading. For instance, although visual metrics (Table~\ref{tab:baseline_comp}) and human inspection (Appendix~\ref{app:fig_comp}) verify the \textit{VAEAttack}'s inferior quality, geometric methods deem it comparable with \textit{UnMarker}.

\noindent
\textit{\textbf{Basic Manipulations.}} We evaluate \textit{Center Cropping ($10\%$)}, \textit{JPEG}, \textit{Quantization}, and \textit{Gaussian Blurring} (filtering). Their parameters (see Appendix~\ref{app:baselines}) were chosen following previous work~\cite{ptw,provable} and to allow maximum distortion while retaining acceptable quality. We also evaluate \textit{Guided Blurring}, showing how even these stronger filters cannot break robust watermarks, as argued in~\S\ref{subsec:semantic}. To better explore these advanced filters, we even evaluate them with overly destructive parameters that do not represent viable attacks. 

\noindent
\textit{\textbf{SOTA Attacks.}}
We compare \textit{UnMarker} against \textit{SOTA} watermark removal attacks: The \textit{DiffusionAttack}~\cite{provable,saberi2023robustness,WAVES} and \textit{VAEAttack}~\cite{provable,WAVES}. To our knowledge, these regeneration methods are the only practical (\textit{query-free} and \textit{black-box}) and effective attacks, as adversarial attacks are disqualified for their unrealistic assumptions (see~\S\ref{sec:Background}).

\subsection{Evaluations}
\subsubsection{Watermark Utility}
\label{subsubsec:utility}

We ensure the chosen schemes' utility, following the requirements from~\S\ref{subsection:freq}: \textit{Applicability}, \textit{Stealthiness}, and \textit{Robustness to Manipulations}. While the latter demands withstanding basic manipulations, true robustness is contingent on resisting advanced attacks as well. We separate these cases for consistency with watermark papers that mainly consider basic manipulations. Advanced attacks are in~\S\ref{subsubsec:sota}. We focus on \textit{Applicability} and \textit{Robustness} as we use original implementations of \textit{SOTA} systems whose \textit{Stealthiness} has been shown by their authors.

Visual metrics are excluded here, as evaluating the naive manipulations' output quality is not our objective, knowing that our \textit{basic} settings are quality-preserving~\cite{provable,ptw}. We do not consider rotation manipulations, following Zhao et al.'s~\cite{provable} arguments w.r.t. their impracticality as they can be easily spotted and corrected before invoking the detector. Still, their finding of some robust schemes vulnerable to rotations may seem at odds with our assertions that these watermarks are in the \textit{collective} spectral amplitudes, which should leave them unaffected by rotations. Yet, schemes are generally not \textit{\textbf{explicitly designed}} to construct watermarks in the spectral amplitudes. Rather, this behavior is adopted when trained for robustness (see~\S\ref{sec:Threory}). Failure to expose these systems to a wider set of demands (e.g., resistance to rotations) may result in them mistakenly associating \textit{\textbf{additional}} indicators, such as a specific orientation, with the watermark. As rotations are impractical, this is not an issue.

\begin{table*}
\caption{Performance of watermarking schemes against each stage of \textit{UnMarker}.}
\label{tab:unmarker}
\renewcommand{\arraystretch}{1}
\fontsize{8pt}{7pt}\selectfont
\begin{tabular}{l}
\\
\!\!\!\!\emph{\underline{Note:} Performance without attacks and under \textit{cropping} is included for reference. The complete attack is in the last row (\textit{CHL}). Numbers in parentheses are}\\
\!\!\!\!\emph{the detection thresholds, while Detect is the portion of images meeting them. The best attacks are in bold font. All schemes lose their \textit{Security Advantage}.}
\vspace{2mm}
\end{tabular}

\renewcommand{\arraystretch}{1.}
\fontsize{11pt}{9pt}\selectfont
\setlength{\tabcolsep}{0.5pt}
\resizebox{\textwidth}{1.5cm}{
\begin{tabular}{ll"cc|cc|cc|cc|cc|cc|cc}
\topthickhline
\addlinespace[5pt]
& & \multicolumn{2}{c}{\textit{\textbf{Yu1}} (61\%)} & \multicolumn{2}{c}{\textit{\textbf{Yu2}} (63\%)} & \multicolumn{2}{c}{\textit{\textbf{HiDDeN}} (73\%)} & \multicolumn{2}{c}{\textit{\textbf{PTW}} (70\%)} & \multicolumn{2}{c}{\textit{\textbf{StableSig.}} (69\%)} & \multicolumn{2}{c}{\textit{\textbf{StegaStamp}} (63\%)} & \multicolumn{2}{c}{\textit{\textbf{TRW}} (0.0141)}\\ 
\multicolumn{2}{c"}
{\diaghead(-2,1){aaaaaaaaaaaaaa}%
{\fontsize{13pt}{11pt}\selectfont \textit{Attack}}{\fontsize{13pt}{11pt}\selectfont \textit{Scheme}}} & Bit Acc$\downarrow$ & Detect$\downarrow$ & Bit Acc$\downarrow$ & Detect$\downarrow$ & Bit Acc$\downarrow$ & Detect$\downarrow$ & Bit Acc$\downarrow$ & Detect$\downarrow$ & Bit Acc$\downarrow$ & Detect$\downarrow$ & Bit Acc$\downarrow$ & Detect$\downarrow$ & Inv. Dist.$\downarrow$ & Detect$\downarrow$\\ 
\midrule
\textbf{\textit{None}} & & 98.42\% & 100\% & 99.71\% & 100\% & 99.16\% & 100\% & 97.96\% & 100\% & 98.97\% & 100\% & 99.95\% & 100\% & 0.022 & 100\% \\ \midrule
\textit{\textbf{Cropping}} & \;(10\%) & 74.56\% & 100\% & 53.71\% & \textbf{0\%} & 86.84\% & 100\% & 93.9\% & 100\% & 97.86\% & 100\% & 93.96\% & 100\% & 0.0155 & 100\% \\ \midrule
\multirow{4}{*}{\bf \textbf{\textit{UnMarker}}} & \multicolumn{1}{c"}{{\large \textit{CH}}} & 59.04\% & 33\% &  53.06\% & \textbf{0\%} &  58.34\% & 1\% & 63.11\% & 11\% &  53.08\% & \textbf{4\%} &  94.87\% & 100\% & 0.0152 & 99\% \\
& \multicolumn{1}{c"}{{\large \textit{CL}}} & 73.16\% & 100\% & 54.62\% & \textbf{0\%} & 84.21\% & 98\% & 95.21\% & 100\% & 97.6\% & 100\% & 61.8\% & \textbf{43\%} & 0.0141 & 55\% \\
& \multicolumn{1}{c"}{{\large \textit{HL}}} & 73.11\% & 100\% & 99.23\% & 100\% & 72.28\% & 66\% & 61.53\% & \textbf{6\%} & 54.59\% & 9\% & 64.64\% & 59\% & 0.016 & 98\% \\ 
& \multicolumn{1}{c"}{{\large \textit{CHL}}} & 59.08\% & \textbf{31\%} & 53.03\% & \textbf{0\%} & 58\% & \textbf{0\%} & 62.75\% & 11\% & 52.82\% & \textbf{4\%} & 61.49\% & \textbf{43\%} & 0.014 & \textbf{40\%} \\ \thickhline
\addlinespace[-5pt]
\end{tabular}
}
\end{table*}
A detailed analysis is in Appendix~\ref{app:baselines} alongside the results (Table~\ref{tab:manipulations}). All schemes detect watermarked images without manipulations: For binary watermarks, all \textit{bit accuracies} approach $100\%$. For \textit{TRW}, the \textit{inverse distance} is $0.022$, well over the $0.0141$ threshold. All systems are fairly resistant to manipulations. \textit{Semantic} schemes remain almost unaffected. Among \textit{non-semantic} schemes, \textit{Yu1} and \textit{StableSignature} are the most resistant, retaining robustness against all \textit{realistic} configurations. The latter shows some vulnerability to \textit{Guided Blurring} with overly destructive parameters, but since this is not a practical attack, it remains reliable. \textit{PTW} rejects most manipulations but is affected by reasonable \textit{Quantization} ($10$), while the less advanced \textit{HiDDeN} is vulnerable to \textit{JPEG} and \textit{Blurring}. \textit{Yu2} resists manipulations except \textit{cropping}, indicating high dependency on spatial attributes, making it unsuitable for watermarking, as it violates the \textit{robustness} requirement from~\S\ref{subsection:freq}.

The results demonstrate how all schemes except \textit{Yu2} are robust to spatial manipulations (e.g., \textit{cropping}), indicating their watermarks are in a different channel, which we next verify is the spectral amplitudes as expected. This is evident for \textit{PTW} and \textit{HiDDeN} that exhibit vulnerability to \textit{Quantization} and \textit{JPEG} that affect high frequencies. The resistance of other schemes to these manipulations merely implies that their watermarks are harder to remove. Yet, advanced attacks can still break these systems, as shown below. \textit{Yu2} highly depends on information in the background, suggesting its watermarks are not constructed as robust, collective measurements but rather in volatile spatial locations.

\subsubsection{\textit{UnMarker}}
\label{subsec:unmarker_eval}
We now evaluate \textit{UnMarker}'s watermark removal capabilities. In~\S\ref{subsubsec:sota}, we compare it to \textit{SOTA} attacks and assess its ability to preserve visual similarity. 

\noindent
\textit{\textbf{Configuration.}} Our analysis above proves that a conventionally robust scheme (under spatial manipulations) must inject its watermarks in the frequency space. Schemes that are spatial in nature and rely less on spectral watermarking (e.g., \textit{Yu2}) may hence evade \textit{UnMarker}'s spectral optimizations as they are not designed to target these inherently fragile schemes. Yet, equipping \textit{UnMarker} with a simple \textit{cropping} layer suffices to defeat such weaker systems. 
Moreover, as spectral watermarks distribute their modifications throughout the image to minimize visual cues (see~\S\ref{sec:Threory}), \textit{cropping} also can slightly ``weaken'' those by eliminating partial modifications residing in the background. This allows using a decreased distortion budget for the optimization stages.

Thus, \textit{UnMarker}'s design is as follows: We apply mild \textit{cropping}, followed by our high-frequency disruptions (\S\ref{subsec:high_freq}). Finally, our filtering stage (\S\ref{subsec:semantic}) eliminates low-frequency traces. This yields a full attack that targets all \textit{carrier}s, erasing all watermarks. Attackers can adjust the \textit{cropping} ratio or location to restrict the output to the actual content. For our experiments, $10\%$ \textit{center cropping} sufficiently retains the content. The stages' order does not affect the results. 

\textit{UnMarker}'s hyperparameters should be calibrated to maximize the spectral loss regardless of the system under attack, complying with our threat model. Hence, they are only affected by properties of the image, such as its size, which we found to be the only relevant factor. The hyperparameters are in Appendix~\ref{app:hyper}. The size-dependant parameters are merely the thresholds and coefficients. Meanwhile, the number of filters and their kernels are invariant, as by combining multiple filter sizes, a single architecture can defeat any \textit{semantic} watermark (see~\S\ref{subsec:semantic}).

\noindent
\textit{\textbf{Results.}} We report settings combining different stages of \textit{UnMarker}, identifying crucial insights that establish our theory. These configurations are \textit{CH}--- the high-frequency destructive process preceded by \textit{cropping}, \textit{CL}--- the low-frequency optimization (with \textit{cropping}), \textit{HL}--- the two optimizations without \textit{cropping}, and \textit{CHL}, denoting the full \textit{UnMarker} with all three stages. Results are in Table~\ref{tab:unmarker}.
Following~\S\ref{subsec:setup}, we say that an attack defeats a scheme (alternatively, \textit{removes its watermark}) if it causes it to lose its \textit{Security Advantage} (i.e., detection below $50\%$).

\noindent
\textit{\underline{Non-Semantic Schemes.}} \textit{Robust} \textit{non-semantic} schemes (w.r.t. known manipulations) cannot alter the content and thus must embed their watermarks in high frequencies. Hence, minimal disruptions to this part of the spectrum where the energy concentration is low suffices to remove them. Our results corroborate this as all such schemes break under \textit{CH} that combines \textit{cropping} and high-frequency disruptions. The detection rates of the \textit{conventionally} robust algorithms (that resist \textit{cropping}--- all but \textit{Yu2}) remain almost unaffected by \textit{CL}, which primarily targets low frequencies where they do not operate. \textit{Yu2} is defeated by \textit{CH}, but \textit{HL}'s results prove this is not due to our optimization since the scheme is insensitive to manipulations in both parts of the spectrum, indicating that \textit{cropping} only is responsible for this outcome. This comes as no surprise since \textit{Yu2} employs \textit{spatial} watermarking that enables it to resist spectral optimizations but makes it vulnerable to \textit{cropping}.

Schemes that process larger images (\textit{PTW} and \textit{StableSignature}) are hardly sensitive to \textit{cropping} but vulnerable to high-frequency manipulations, shown by their robustness to \textit{CL} and fragility against \textit{HL}, bringing their detection to $6\%$ and $9\%$. \textit{Yu1} and \textit{HiDDeN} that watermark smaller {\small \mbox{$128 {\times} 128$}} images retain $100\%$ detection against \textit{HL}. Yet, their \textit{bit accuracies} drop sharply from $98.42\%$ and $99.16\%$ to $73.11\%$ and $72.28\%$. The cause is our high-frequency destruction since \textit{CL} that combines \textit{cropping} and low-frequency disruptions offers almost no advantages over \textit{cropping}.

Thus, there is a connection between the image size, sensitivity to \textit{cropping}, and vulnerability to high-frequency modifications, which is easy to explain: Spatial \textit{robustness} and \textit{stealthiness} require spreading the watermark across the image. Meanwhile, \textit{non-semantic} watermarks are restricted to specific regions (i.e., edges) where they can exploit the high-frequency nature (see~\S\ref{subsubsec:candidates}) to remain invisible. As the size shrinks, such areas become less prevalent in central parts, forcing the changes to be delegated to the background and making them more vulnerable to \textit{cropping}. Accordingly, the portion of the change each region receives grows as well, making it difficult to eliminate these watermarks with minimal modifications, explaining why they retain \textit{bit accuracies} above the threshold despite a significant decline without \textit{cropping}. Regardless, \textit{mild cropping} then becomes capable of eliminating sufficient footprints to further decrease these values below the thresholds, as seen for \textit{CH} (\textit{CHL}).

Larger images enable schemes to better reject \textit{cropping} since this new-found resolution facilitates embedding the watermark in central areas, but the footprint in these regions must be smaller. Thus, minute disruptions throughout the image become more effective, explaining their inability to withstand high-frequency optimizations alone.

\textbf{\textit{UnMarker} (\textit{CHL}) defeats all \textit{non-semantic} schemes, with the most resistant being \textit{Yu1}, which retains a detection rate of $31\%$. Yet, Yu1 still loses its Security Advantage (and so do all other schemes).}

\noindent
\textit{\underline{Semantic Schemes.}} \textit{StegaStamp} and \textit{TRW} resist high-frequency disruptions (\textit{CH}), with their detection fixed at $100\%$ and the \textit{bit accuracies} (\textit{inverse distances}) retaining almost identical values to those observed for \textit{cropping} alone. Yet, both are vulnerable to low-frequency disruptions, evident by their significant performance decline against \textit{CL}: \textit{StegaStamp}'s detection drops to $43\%$ while \textit{TRW}'s becomes $55\%$. This cements our arguments regarding \textit{semantic} watermarks being primarily in lower frequencies and the soundness of our filtering approach. Comparing \textit{CL} to the final results (\textit{CHL}), we observe identical detection for \textit{StegaStamp}. \textit{TRW} exhibits an additional decline to $40\%$, indicating a small portion of its watermark is in high frequencies.

\textit{StegaStamp}'s watermarks are more visible (see~\S\ref{subsec:defensive}), better engraving them in the \textit{main} content, and increasing their robustness to spatial manipulations (\textit{cropping}). Since \textit{TRW} produces less visible watermarks despite targeting low frequencies where significant watermarking footprints are required, it behaves similar to \textit{non-semantic} schemes for small images, forcing a considerable portion of these changes to populate the background. The results for \textit{cropping}, \textit{CL, CH} and \textit{CHL} prove this; while \textit{cropping} minimally affects \textit{StegaStamp} (the \textit{bit accuracy} only drops to $93.96\%$), \textit{TRW} is more sensitive as its \textit{inverse distance} decreases from $0.022$ to $0.155$ when the threshold is only $0.0141$. Although this is in the detection range, combining \textit{cropping} with \textit{UnMarker}'s low-frequency disruptions (\textit{CL, CHL}) erases its watermarks (eliminates its \textit{Security Advantage}). 

Spectral optimizations (\textit{HL}) drastically lower \textit{TRW}'s \textit{inverse distance} (to $0.016$), but do not erase its watermarks. This is because low-frequency distortions affect many pixels simultaneously. Under a strict budget, the number of regions where these rates can be altered is limited compared to high-frequencies. \textit{StegaStamp}, whose watermarks are centralized, requires modifying fewer areas, making \textit{HL} successful without \textit{cropping}. Yet, we verified that low-frequency disruptions with a higher budget or number of steps break \textit{TRW} without \textit{cropping}, but this may introduce artifacts unless carefully done. \textit{Cropping} is as effective but less demanding, confirming a flaw in \textit{TRW} that targets non-critical regions.

\begin{table*}[ht]
\caption{Comparison with \textit{SOTA} watermark removal attacks.
}
\label{tab:baseline_comp}

\renewcommand{\arraystretch}{1}
\fontsize{8pt}{8pt}\selectfont
\begin{tabular}{l}
\!\!\!\!\emph{\underline{Note}: Performance in the absence of attacks is included for reference. For each metric, bold entries correspond to the best attack against the relevant scheme.} \\

\!\!\!\!\emph{Gray cells denote scores meeting our quality and removal criteria for \textit{LPIPS}--- 0.15 and Detect--- $50\%$ (where the system} loses its \textit{Security} \textit{Advantage}).
\end{tabular}

\vspace{2mm}

\renewcommand{\arraystretch}{1}
\fontsize{8pt}{8pt}\selectfont
\begin{tabular}{llllll}
\multicolumn{4}{@{}l}{\em(a) \fontsize{11pt}{9pt}\selectfont Performance against \textit{non-semantic} schemes.}\\
\end{tabular}

\renewcommand{\arraystretch}{7.5}
\fontsize{190pt}{21pt}\selectfont
\resizebox{\textwidth}{1.6cm}{
\begin{tabular}{ll@{\hspace{-5ex}}"@{\hspace{1ex}}cccc|cccc|cccc|cccc|cccc}

\toprule
\addlinespace[40pt]
& & \multicolumn{4}{c}{\textit{\textbf{Yu1}} (61\%)} & \multicolumn{4}{c}{\textit{\textbf{Yu2}} (63\%)} & \multicolumn{4}{c}{\textit{\textbf{HiDDeN}} (73\%)} & \multicolumn{4}{c}{\textit{\textbf{PTW}} (70\%)} & \multicolumn{4}{c}{\textit{\textbf{StableSignature}} (69\%)} \\ 
\multicolumn{2}{l"@{\hspace{1ex}}}
{\diaghead(-6,1){aaaaaaaaaaaaaaaaaaaaaa}%
{\fontsize{250pt}{9pt}\selectfont \textit{Attack}}{\fontsize{250pt}{9pt}\selectfont \textit{Scheme}}} & \textit{FID}$\downarrow$ & \textit{LPIPS}$\downarrow$ & Bit Acc$\downarrow$ & Detect$\downarrow$ & \textit{FID}$\downarrow$ & \textit{LPIPS}$\downarrow$ & Bit Acc$\downarrow$ & Detect$\downarrow$ & \textit{FID}$\downarrow$ & \textit{LPIPS}$\downarrow$ & Bit Acc$\downarrow$ & Detect$\downarrow$ & \textit{FID}$\downarrow$ & \textit{LPIPS}$\downarrow$ & Bit Acc$\downarrow$ & Detect$\downarrow$ & \textit{FID}$\downarrow$ & \textit{LPIPS}$\downarrow$ & Bit Acc$\downarrow$ & Detect$\downarrow$ \\ [-10pt]
\midrule
\addlinespace[20pt]
{\fontsize{200pt}
{10pt}\selectfont
\textbf{\textit{None}}} & & \textit{NA} & \textit{NA} & 98.42\% & 100\% & \textit{NA} & \textit{NA} & 99.71\% & 100\% & \textit{NA} & \textit{NA} & 99.16\% & 100\% & \textit{NA} & \textit{NA} & 97.96\% & 100\% & \textit{NA} & \textit{NA} & 98.97\% & 100\% \\ \thickhline \midrule
\addlinespace[20pt]
{\fontsize{200pt}{10pt}\selectfont
\textbf{\textit{UnMarker (CHL)}}} & & 34.07 & \cellcolor[gray]{0.9} \textbf{0.15} & 59.08\% & \cellcolor[gray]{0.9} 31\% & 22.88 & \cellcolor[gray]{0.9} \textbf{0.1} & 53.03\% & \cellcolor[gray]{0.9} \textbf{0\%} & 39.56 & \cellcolor[gray]{0.9} \textbf{0.08} & 58\% & \cellcolor[gray]{0.9} \textbf{0\%} & \textbf{11.9} & \cellcolor[gray]{0.9} \textbf{0.14} & 62.75\% & \cellcolor[gray]{0.9} 11\% & \textbf{30.87} & \cellcolor[gray]{0.9} \textbf{0.05} & 52.82\% & \cellcolor[gray]{0.9} 4\% \\ \thickhline
\addlinespace[20pt]
{\fontsize{200pt}{10pt}\selectfont\multirow{2}{*}{\begin{tabular}{c}
\textbf{\textit{Diffusion}}  \\
\textbf{\textit{Attack}}
\end{tabular}}} & {\fontsize{190}{40}\selectfont \!\!\!\!\!\!\!\!\!\!\!\!\!\!\!\!\!\!\!\textit{no cropping}} & \textbf{12.38} & 0.2 & 58.31\% & \cellcolor[gray]{0.9} 28\% & 13.09 & 0.17 & 68.29\% & 87\% & \textbf{34.1} & 0.18 & 59.8\% & \cellcolor[gray]{0.9} 3\% & 22.95 & 0.26 & \textbf{55.53\%} & \cellcolor[gray]{0.9} \textbf{7\%} & 36.57 & 0.32 & 47.89\% & \cellcolor[gray]{0.9} \textbf{0\%} \\
\addlinespace[20pt]
& {\fontsize{190}{40}\selectfont \!\!\!\!\!\!\!\!\!\!\!\!\!\!\!\!\!\!\!\textit{+cropping}} & 14.02 & 0.18 & \textbf{54.24\%} & \cellcolor[gray]{0.9} \textbf{8\%} & \textbf{11.98} & 0.17 & \textbf{51.26\%} & \cellcolor[gray]{0.9} 3\% & 36.97 & 0.19 & 56.61\% & \cellcolor[gray]{0.9} \textbf{0\%} & 19.98 & 0.25 & 62.01\% & \cellcolor[gray]{0.9} 16\% & 41.88 & 0.31 & 47.78\% &\cellcolor[gray]{0.9}  \textbf{0\%} \\ [20pt]
\thickhline
\addlinespace[20pt]
{\fontsize{200pt}{10pt}\selectfont\multirow{3}{*}{\begin{tabular}{c}
\textbf{\textit{VAE}}  \\
\textbf{\textit{Attack}}
\end{tabular}}} & {\fontsize{190}{77}\selectfont \!\!\!\!\!\!\!\!\!\!\!\!\!\!\!\!\!\!\!\textit{quality=3}} & 36.2 & \cellcolor[gray]{0.9} \textbf{0.15} & 67.19\% & 97\% & 19.67 & \cellcolor[gray]{0.9} 0.13 & 73.89\% & 96\% & 65.16 & 0.22 & 60.4\% & \cellcolor[gray]{0.9} 8\% & 26.6 & 0.26 & 73.63\% & 89\% & 72.21 & 0.37 & 48.25\% & \cellcolor[gray]{0.9} \textbf{0\%} \\
\addlinespace[20pt]
& {\fontsize{190}{40}\selectfont \!\!\!\!\!\!\!\!\!\!\!\!\!\!\!\!\!\!\!\textit{quality=2}} & 46.82 & 0.18 & 64.42\% & 81\% & 27.4 & 0.16 & 66.62\% & 78\% & 83.4 & 0.26 & 58.36\% & \cellcolor[gray]{0.9} 1\% & 39.11 & 0.29 & 71.33\% & 79\% & 94.85 & 0.41 & 44.93\% & \cellcolor[gray]{0.9} \textbf{0\%} \\
\addlinespace[20pt]
& {\fontsize{190}{40}\selectfont \!\!\!\!\!\!\!\!\!\!\!\!\!\!\!\!\!\!\!\textit{quality=1}} & 58.37 & 0.22 & 60.63\% & 59\% & 38.9 & 0.19 & 61\% & \cellcolor[gray]{0.9} 36\% & 104.92 & 0.32 & \textbf{56.26\%} & \cellcolor[gray]{0.9} 2\% & 54.93 & 0.33 & 66.74\% & \cellcolor[gray]{0.9} 36\% & 128.88 & 0.44 & \textbf{43.87\%} & \cellcolor[gray]{0.9} \textbf{0\%} \\ 
\bottomrule
\\
\end{tabular}
}
\end{table*}
\begin{table}
\vspace{-5mm}
\renewcommand{\arraystretch}{1}
\fontsize{9pt}{9pt}\selectfont
\begin{tabular}{llllll}
\multicolumn{4}{@{}l}{\em(b) \fontsize{11pt}{9pt}\selectfont Performance against \textit{semantic} schemes.}\\
\end{tabular}
\renewcommand{\arraystretch}{7.5}
\fontsize{190pt}{21pt}\selectfont

\resizebox{\columnwidth}{1.6cm}{
\begin{tabular}{ll@{\hspace{-5.5ex}}"@{\hspace{1ex}}cccc|cccc}

\toprule
\addlinespace[40pt]
& & \multicolumn{4}{c}{\textit{\textbf{StegaStamp}} (63\%)} & \multicolumn{4}{c}{\textit{\textbf{TRW}} (0.0141)}\\
\multicolumn{2}{l"@{\hspace{1ex}}}{\diaghead(-6,1){aaaaaaaaaaaaaaaaaaaaaaaa}%
{\fontsize{250pt}{9pt}\selectfont \textit{Attack}}{\fontsize{250pt}{9pt}\selectfont \textit{Scheme}}} 
 & \textit{FID}$\downarrow$ & \textit{LPIPS}$\downarrow$ & Bit Acc$\downarrow$ & Detect$\downarrow$ & \textit{FID}$\downarrow$ & \textit{LPIPS}$\downarrow$ & Inv. Dist.$\downarrow$ & Detect$\downarrow$ \\ [-10pt]
\midrule
\addlinespace[20pt]
{\fontsize{200pt}
{10pt}\selectfont
\textbf{\textit{None}}} & & \textit{NA} & \textit{NA} & 99.95\% & 100\% & \textit{NA} & \textit{NA} & 0.022 & 100\% \\ \thickhline \midrule
\addlinespace[20pt]
{\fontsize{200pt}{10pt}\selectfont
\textbf{\textit{UnMarker (CHL)}}} & & 34.69 & \cellcolor[gray]{0.9} \textbf{0.08} & \textbf{61.49\%} & \cellcolor[gray]{0.9} \textbf{43\%} & 60.73 & \cellcolor[gray]{0.9} \textbf{0.1} & \textbf{0.014} & \cellcolor[gray]{0.9} \textbf{40\%}\\ \thickhline
\addlinespace[20pt]
{\fontsize{200pt}{10pt}\selectfont\multirow{2}{*}{\begin{tabular}{c}
\textbf{\textit{Diffusion}}  \\
\textbf{\textit{Attack}}
\end{tabular}}} & {\fontsize{190}{40}\selectfont \!\!\!\!\!\!\!\!\!\!\!\!\!\!\!\!\!\!\!\textit{no cropping}} & 33.15 & 0.36 & 83.21\% & 100\% & 41.69 & 0.32 & 0.0155 & 99\% \\
\addlinespace[20pt]
& {\fontsize{190}{40}\selectfont \!\!\!\!\!\!\!\!\!\!\!\!\!\!\!\!\!\!\!\textit{+cropping}} & \textbf{30.83} & 0.33 & 77.84\% & 95\% & \textbf{41.16} & 0.31 & 0.0144 & 79\% \\ [20pt]
\thickhline
\addlinespace[20pt]
{\fontsize{200pt}{10pt}\selectfont\multirow{3}{*}{\begin{tabular}{c}
\textbf{\textit{VAE}}  \\
\textbf{\textit{Attack}}
\end{tabular}}} & {\fontsize{190}{40}\selectfont \!\!\!\!\!\!\!\!\!\!\!\!\!\!\!\!\!\!\!\textit{quality=3}} & 54.89 & 0.42 & 97.9\% & 100\% & 71.91 & 0.38 & 0.0154 & 90\% \\
\addlinespace[20pt]
& {\fontsize{190}{40}\selectfont \!\!\!\!\!\!\!\!\!\!\!\!\!\!\!\!\!\!\!\textit{quality=2}} & 70.86 & 0.46 & 94.7\% & 100\% & 93.56 & 0.41 & 0.015 & 88\% \\
\addlinespace[20pt]
& {\fontsize{190}{40}\selectfont \!\!\!\!\!\!\!\!\!\!\!\!\!\!\!\!\!\!\!\textit{quality=1}} & 87.8 & 0.49 & 88.84\% & 100\% & 118.85 & 0.46 & 0.0145 & 67\% \\ 
\bottomrule
\end{tabular}}
\end{table}

\textbf{
\textit{UnMarker} defeats all semantic schemes due to irreconcilable trade-offs, as robustness to spatial manipulations incurs vulnerability to spectral disruptions and vice versa. Even when not bringing detection to $0\%$, it leaves a scheme useless, eliminating its \textit{Security Advantage}}.

\noindent
\underline{\textit{Remarks:}} \textbf{Any} scheme either modifies the content (\textit{semantic}) or does not (\textit{non-semantic}). If attackers know the scheme's class, they can apply only one of the stages (\textit{CH} or \textit{CL}), sufficing to erase the watermark, avoiding additional distortions. This information may be public or obtained by attacking multiple fake images (not the one intended for use in the actual attack), using the two stages independently, and later observing which images were not deleted (flagged). Some schemes may use principles from both classes. For instance, although \textit{TRW} is \textit{semantic}, a portion of its watermarks is in high frequencies, as discussed above. Here, applying both stages boosts the success rates. Yet, for all schemes (including less sophisticated ones, where only one stage is required), the full \textit{UnMarker} (\textit{CHL}) preserves quality even when both stages are applied (see~\S\ref{subsubsec:sota}).

\subsubsection{Comparison with \textit{SOTA} attacks}
\label{subsubsec:sota}
We compare \textit{UnMarker} to regeneration attacks--- the \textit{VAEAttack}~\cite{provable,WAVES} and \textit{DiffusionAttack}~\cite{provable,saberi2023robustness,WAVES}. Despite being certifiably guaranteed to remove \textit{non-semantic} watermarks, they fail against \textit{semantic} methods. This robustness creates a false sense of security around \textit{semantic} watermarks~\cite{provable,WAVES}, which \textit{UnMarker} disproves. Additionally, they guarantee the removal of \textit{non-semantic} watermarks only if the attack introduces sufficient noise to dilute the watermark, followed by a denoising process to restore the image's quality. Stronger watermarks require increased perturbations that prevent perfect reconstruction. Unnatural artifacts can be avoided by using advanced regeneration systems, explaining why the more sophisticated \textit{DM}s outperform \textit{VAE}s, as they can apply larger distortions while somewhat accurately retrieving the content. Yet, similarity to the original image may still be lost under such large budgets. \textit{UnMarker} eliminates these hurdles, significantly decreasing the distortion while being effective even against the stronger \textit{semantic} schemes.

As \textit{UnMarker} uses \textit{cropping}, we apply the \textit{DiffusionAttack} with and without \textit{cropping} for a fair comparison. The \textit{VAEAttack}'s image quality was deemed inferior, disqualifying many of its results (see below), and thus a similar comparison was excluded. Results are in Table~\ref{tab:baseline_comp}, and samples produced by the three attacks are in Appendix~\ref{app:fig_comp}.

\noindent
\textit{\underline{Quality:}} The range for \textit{LPIPS} is $[0, 1]$. Images with \textit{LPIPS} distance at the $0.1$ level (i.e., below $0.2$) are visually similar~\cite{hao2021s}. Thus, we choose $0.15$ as a cut-off beyond which we say the attack begins to cause visible changes. This does not mean that slightly higher scores imply severe degradation since the changes may integrate well with the content. Yet, scores far beyond this threshold attest to poor output nature. Despite \textit{FID}'s known ability to measure quality, it has two major shortcomings: first, it compares the \textit{datasets} of watermarked images and those that have undergone the attack and cannot judge samples on an image-by-image basis. It may also exhibit inconsistencies with human vision, making it less ideal in certain instances~\cite{liu2018improved}. Thus, while \textit{FID} is important for assessing quality, we prioritize \textit{LPIPS}.

We do not consider the \textit{VAEAttack} effective. It allows choosing from eight quality options that control the added noise before regeneration. We apply it with the three lowest options, as we found increasing this factor beyond $3$ makes the attack almost always ineffective. Yet, all three options severely impact the quality, invalidating the attack. This can be seen in Table~\ref{tab:baseline_comp}, as both its \textit{FID} and \textit{LPIPS} are extremely worse (larger) than the other attacks whenever it degrades the detection rates. For larger images (over ${\text{\small \mbox{$256 {\times} 256$}}}$), even the best factor ($3$) is destructive, resulting in \textit{LPIPS} scores at least $3.8\times$ compared to \textit{UnMarker} and well above the $0.15$ cut-off. Its \textit{FID}s are also inflated. While its \textit{FID} against \textit{TRW} ($71.91$) is close \textit{UnMarker}'s ($60.73$), its \textit{LPIPS} is much worse, indicating low quality. For smaller images, it retains similarity for \textit{Yu1, Yu2} and \textit{PTW} with $quality\!\!=\!\!3$, but fails to remove their watermarks. \textit{HiDDeN}'s watermarks are removed with this factor at an acceptable \textit{LPIPS} ($0.22$). Still, the failure against other systems limits this attack. Lowering the factor impacts quality, invalidating any results for larger images (including \textit{PTW}). For \textit{Yu1} and \textit{Yu2}, as the attack fails to achieve remarkable removal for $quality\!\!=\!\!2$, the relevant case is $quality\!\!=\!\!1$. Both systems retain acceptable \textit{LPIPS} scores ($0.22$ and $0.19$) but violate the $0.15$ threshold, and their \textit{FID}s are far worse than the other attacks, proving the \textit{VAEAttack} fails to preserve quality (see Appendix~\ref{app:fig_comp}).

The \textit{DiffusionAttack} is controlled via a parameter~$t$. Following Saberi et al.~\cite{saberi2023robustness}, $t\!>\!0.2$ causes significant deviations from the watermarked input, invalidating the attack. We found that for smaller images ({\small \mbox{$128 {\times} 128$}}), this occurs at lower values. For these images, we set $t\!\!=\!\!0.05$, deeming it the turning point. A Lower $t$ results in less successful attacks. \textit{UnMarker} achieves better \textit{LPIPS} ($\leq\!\!0.15$), ensuring similarity. The differences are drastic for larger images (over ${\text{\small \mbox{$256 {\times} 256$}}}$)--- $0.1$ at most for \textit{UnMarker} compared to $\geq\!0.37$ for the \textit{DiffusionAttack}. The higher \textit{LPIPS} of \textit{DiffusionAttack}, while not necessarily indicating significant deviations, entail that some nuances will appear, which may lead to its outputs being identified as fake--- see Appendix~\ref{app:fig_comp}. The attacks' \textit{FID}s alternate, occasionally outperforming one another. Yet, both attain far better \textit{FID}s than the \textit{VAEAttack}.

\textbf{
\textit{UnMarker} attains the lowest \textit{LPIPS}, with its worst being $0.15$, ensuring resemblance to the watermarked images. The similarity between \textit{UnMarker}'s \textit{FID}s and the \textit{DiffusionAttack}'s, which is known to produce high-quality outputs~\cite{provable}, further establishes its capabilities.
}

\noindent
\textit{\underline{Watermark Removal:}}
Both \textit{SOTA} attacks defeat \textit{non-semantic} schemes, although the \textit{VAEAttack}'s quality invalidates its results. \textit{UnMarker} always outperforms this attack except against \textit{StableSignature} for which \textit{UnMarker} brings the detection to $4\%$ while the \textit{VAEAttack} reduces it to $0\%$. Still, this difference is insignificant. The \textit{DiffusionAttack} and \textit{UnMarker} have similar results on average, but the \textit{DiffusionAttack} (+\textit{cropping}) significantly outperforms \textit{UnMarker} on \textit{Yu1} ($8\%$ detection vs. $31\%$). Yet, as both eliminate its \textit{Security Advantage}, specific numbers are irrelevant.

Our results validate regeneration's known inability to remove \textit{semantic} watermarks~\cite{provable,saberi2023robustness,WAVES}. The \textit{VAEAttack} moderately affects \textit{TRW}, especially with quality $1$, bringing its detection to $67\%$, but, as explained above, we consider it invalid. The \textit{DiffusionAttack} leaves both detection rates almost unchanged despite reducing \textit{StegaStamp}'s \textit{bit accuracy} and \textit{TRW}'s \textit{inverse distance} from $99.95\%$ and $0.022$ to $83.21\%$ and $0.0155$. While \textit{cropping} decreases these numbers to $77.84\%$ and $0.0144$ and makes the attack more successful against \textit{TRW} ($79\%$ detection), \textit{UnMarker} drastically outperforms it, reducing this to $40\%$ and \textit{StegaStamp}'s, against which \textit{DiffusionAttack} is ineffective, to $43\%$.

\textbf{Among the \textit{SOTA} attacks, only the \textit{DiffusionAttack} can remove watermarks without violating visual constraints. Yet, \textit{UnMarker} is superior as it succeeds against \textit{semantic} watermarks where the \textit{DiffusionAttack} fails}.

%% file: Sections/discussion.tex
\section{Discussion}
\label{section:discussion}
\noindent
\textit{\textbf{Practicality.}} \textit{UnMarker} is executed offline as it requires no detector interaction and takes up to $5$ minutes with a \textit{$40GB$ A100 GPU}. Such machines are available for $\$32.77/hr$ on \textit{AWS}. It requires no data, freeing it from the dependence on denoising pipelines (e.g., \textit{DM}s). Regeneration attacks are affected by these systems' fidelity and the commonalities between their training data and the watermarked images, as we showed for the \textit{VAEAttack}. While the \textit{DiffusionAttack} achieves better quality (but worse than \textit{UnMarker}'s), its \textit{DM} and all studied schemes were trained on public data. Providers possess exclusive data that differ from those used to train the attacker's \textit{DM}, making it likely to impact quality.

\noindent
\textit{\textbf{Mitigations.}} In the context of defensive watermarking, mitigations should allow recovery from removal attacks, enabling providers to continue using the scheme. Adversarial training (\textit{AT})~\cite{adversarial_training} is a popular mitigation in \textit{ML}, wherein the model (detector) is trained on malicious samples from known attacks so that it learns to classify them correctly. Mitigations can also target the model's inputs: the provider applies a pre-processing stage (e.g., purification~\cite{nie2022DiffPure}) to eliminate disruptions and only then invokes the model (detector). Mitigations could exist for adversarial attacks as they \textit{typically} retain the sample's true class but rely on model feedback to exploit its imperfections, yielding ``volatile'' perturbations in its uncertainty regions that cause misclassification. This strategy has also been the focus of previous work on watermark removal~\cite{lukas2023leveraging,WeVade,saberi2023robustness}. Yet, if the watermark is actually removed, recovery would be impossible. As we claim that by simply disrupting the \textit{carrier}, \textit{UnMarker} removes the watermark, mitigations should fail.

To test this, we consider a \textit{SOTA} mitigation--- diffusion purification~\cite{nie2022DiffPure}.  Purification inspires the \textit{DiffusionAttack} (see~\S\ref{subsubsec:sota}), operating similarly: It adds noise, diluting adversarial perturbations, and uses a \textit{DM} for reconstructing a pure (attack-free) sample. As this applies as a certifiable attack on \textit{non-semantic} watermarks, these schemes cannot use this mitigation. Yet, these watermarks are of little importance since these proofs for their removal undermine their practicality. As we aim to investigate the validity of defensive watermarking, we focus on the more reliable \textit{semantic} schemes that meet the standards for utilizing purification as a defense. Still, \textit{UnMarker}'s stages against both classes hinge on similar principles, making the findings applicable to defenses for \textit{non-semantic} schemes if implemented. 

We attack $100$ samples for each scheme using \textit{UnMarker} (\textit{CHL}). Then, we perform purification and invoke the detector. Given the resistance of \textit{semantic} schemes to purification (the \textit{DiffusionAttack}) and the effectiveness of this mitigation against adversarial examples, if \textit{UnMarker} does not remove the watermark but overshadows it with volatile changes, purification would eliminate \textit{UnMarker} and restore the watermark. Yet, the results verify \textit{UnMarker}'s destructive nature that makes recovery impossible. For purification with $t\!\!=\!\!0.1$ (see~\S\ref{subsubsec:sota}), \textit{StegaStamp}'s detection drops to $15\%$ and \textit{TRW}'s to $16\%$. A larger $t$ leads to worse detection. Ironically, the defense degrades performance. While this is partly because diffusion may slightly decrease the \textit{bit accuracy} (\textit{inverse distance}), we apply it at half of its original strength (see~\S\ref{subsubsec:sota}), which allowed for $\sim \!\!100\%$ detection. The reason for this decline is \textit{UnMarker}'s destruction, impairing the \textit{DM}'s reconstruction ability. This raises the possibility of combining \textit{UnMarker} with the \textit{DiffusionAttack} as an even stronger attack, which we leave to future work due to the potential increased quality degradation. Evaluating \textit{AT} is prohibitive due to time limitations, as it requires numerous adversarial examples, and a single \textit{UnMarker} attack takes several minutes. Yet, \textit{AT} is only successful if the attack sample retains the watermark, which is not the case for \textit{UnMarker}, as proven. As purification adheres to similar principles, its ineffectiveness implies the same for \textit{AT}.

\noindent
\textit{\textbf{Outlook.}} A natural question is whether watermarking can ever be made reliable. Particularly, can we devise schemes that do not conform to our spectral analysis s.t. \textit{UnMarker}, or similar attacks, would fail? The answer is likely no. 

In~\S\ref{sec:Threory}, we establish that a \textit{universal carrier} that encodes the watermark is crucial for any scheme due to computational considerations. Given the nature of visual content, the assertion that this \textit{carrier} must be in the spatial structure or spectrally constructed is also invariant. Then, we explain how spatial dependencies lead to fragile watermarks, as is demonstrated in~\S\ref{subsection:freq}. Thus, robust schemes must embed watermarks in the \textit{collective} spectral amplitudes. \textit{Non-semantic} schemes are unreliable and, despite their slightly demanding nature, regeneration attacks \textit{certifiably} erase them, restricting strong watermarks to lower frequencies (\textit{semantic}). While these are of larger magnitudes, they must remain minimally visible to retain quality and since obvious traces are removable even via naive editing. Existing \textit{semantic} schemes push these demands to the extreme, leaving somewhat visible footprints (\textit{StegaStamp}) or changing the content and limiting the space of watermarked images (\textit{TRW}). Since \textit{UnMarker} defeats them, future variants will have to further influence the content. Due to the considerable modifications the \textit{SOTA} methods incur, such schemes will likely be impractical, reducing the quality and variety of watermarkable content and leaving visible traces. Combined with the inefficacy of mitigations, defensive watermarking becomes invalid, and future work should explore alternatives. 

\noindent
\textit{\textbf{Limitations \& Future Work.}} 1) We focus on practical (current) systems applicable to various images. Future schemes might be tailored for specific content (e.g., adding brand-specific glasses to face images), but such \textit{GenAI} services will lack diversity. To meet users’ needs, providers will also offer less restrictive tools, and attackers can use them instead, rendering content-specific watermarking useless. To reduce visibility and resist naive attacks, they must also employ spectral manipulations (for reasons similar to~\S\ref{subsubsec:candidates}'s), leaving \textit{UnMarker} effective. Yet, without requiring universal applicability, the spectral footprint may target certain regions or objects only, not the collective spectrum. This necessitates augmenting \textit{UnMarker} with content-based knowledge to identify optimal disruption regions, which could be studied should such schemes become prevalent. 2) \textit{UnMarker} relies on accurate visual losses that preserve essential similarity cues while maximizing perturbations to attain the best removal rates, and future work may explore alternatives superior to ours (see~\S\ref{subsec:high_freq}). 3) While we consider visual content, our analysis translates to other domains (e.g., audio), but \textit{UnMarker} requires domain-based adaptations that can be interesting to investigate. 4) As watermarks bear resemblance to adversarial examples (\textit{AE}s), \textit{UnMarker} could potentially defend against \textit{AE}s. Still, as \textit{AE}s do not require a universal carrier, they may target other channels (aside from collective spectra), mandating further improvements.

\noindent
\textit{\textbf{Ethical Considerations.}} Misusing \textit{GenAI} has severe impacts on society. We aim not to help generate harmful content but raise awareness regarding these systems' risks. Although we are not aware of providers that already released stable watermarking schemes, their adoption pace and attention from dominant entities (see~\S\ref{sec:intro}) make assessments mandatory. We requested access to providers' \textit{API}s for evaluation but never heard back. Thus, all studied systems are open-source.

%% file: Sections/Conclusion.tex
\section{Conclusion}
We presented a novel attack against defensive watermarking via adversarial optimizations that disrupt the spectra of watermarked images and interfere with the watermarks' footprints, erasing them without interacting with the scheme. \textit{UnMarker} is effective against all systems, including \textit{semantic} schemes that have been assumed robust to such \textit{query-free} and \textit{black-box} attacks. Our results show that defensive watermarking is not a viable countermeasure against deepfakes and that more robust defenses are needed.

%% file: Sections/ack.tex
\ifCLASSOPTIONcompsoc
  \section*{Acknowledgments}
\else
  \section*{Acknowledgment}
\fi

We gratefully acknowledge the support of the Waterloo-Huawei Joint Innovation Laboratory for funding this research, and the Compute Canada Foundation (CCF) for their resources that made our experiments possible.

%% file: Sections/appendix.tex
\section{\textit{UnMarker}'s Hyperparameters} 
\label{app:hyper}

\noindent
\textit{\underline{High-frequency Disruptions} (Stage 1---~\S\ref{subsec:high_freq})}: We use \textit{Alex}~\cite{lpips} for small images ({\small \mbox{$128 {\times} 128$}}) and \textit{DVL}~\cite{deeplossVGG} for larger sizes as the perceptual loss ($\ell_p$). As we maximize \textit{DFL}, we set $t_{{\ell_p}}$ to the largest value at which similarity and quality are retained. This depends on $\ell_p$ and the image size. For small images optimized using \textit{Alex}, this optimal point is $t_{{{\text{\tiny \mbox{${Alex}$}}}}}\!\!\!=\!\!5 \!\cdot\! 10^{-4}$. Since larger images require a different loss (\textit{DVL}), similar thresholds do not indicate the same degree of resemblance. For {\small \mbox{$256 {\times} 256$}} images, we find $t_{{{\text{\tiny \mbox{${DVL}$}}}}}\!\!=\!4\! \cdot\! 10^{-2}$ best achieves the above goals, while for larger images we have $t_{{{\text{\tiny \mbox{${DVL}$}}}}}\!\!=\!\!10^{-4}$. Next, we select the initial $c_{{\ell_p}}$ that balances the spectral and visual losses. We want this starting value to be significant to assign a large weight to $\ell_p$ and ensure visual similarity is not violated, after which it can be decreased to allow the spectral loss to be more dominant while still ensuring this property. This initial setting only affects the time until it converges and must be empirically determined to speed up the process. We found an initial $c_{{{\text{\tiny \mbox{${Alex}$}}}}}\!\!=\!\!10^6$ optimal for small images. For \textit{DVL}, this was set at $c_{{{\text{\tiny \mbox{${DVL}$}}}}}\!\!=\!\!10^6$ when the size is above {\small \mbox{$256 {\times} 256$}} as these images require the \textit{DVL} loss to be extremely small ($t_{{{\text{\tiny \mbox{${DVL}$}}}}}\!\!=\!\!10^{-4}$), while for {\small \mbox{$256 {\times} 256$}} images, it can be far smaller ($c_{{{\text{\tiny \mbox{${DVL}$}}}}}\!\!=\!\!10^{-2}$).

The $\|\|_2$ component prevents unwanted peaks (see~\S\ref{subsec:high_freq}). Its threshold depends on the size and chosen \textit{LPIPS}. While the \textit{LPIPS} thresholds guarantee visual similarity to the extent that the images will look similar up to, potentially, such peaks, $\|\|_2$'s thresholds must eliminate these singular points when the \textit{LPIPS} thresholds are met. We empirically selected these $\|\|_2$ thresholds: for {\small \mbox{$128 {\times} 128$}}, we have $t_{{\|\|_2}}=10^{-4}$, while {\small \mbox{$256 {\times} 256$}} images allow $t_{{\|\|_2}}=3 \cdot 10^{-5}$, and larger images operate at $t_{{\|\|_2}}=10^{-4}$. We report the normalized thresholds (divided by the image size). In practice, they increase with the size since larger images accommodate stronger perturbations without visible cues. $c_{{\|\|_2}}$ multiplies this $\|\|_2$ when its threshold is violated to strongly steer the procedure away from these boundaries. This is not determined through a binary search, as we want it to have the same effect constantly, without undoing the previous iterations' effects. By tracking the losses, we arrive at $c_{{\|\|_2}}=0.6$.

\begin{table*}
\caption{Performance of watermarking schemes against basic manipulations.}
\label{tab:manipulations}
\large

\renewcommand{\arraystretch}{1}
\fontsize{8pt}{7pt}\selectfont
\begin{tabular}{l}
\!\!\!\!\emph{\underline{Note:} The first row shows the results in the absence of any manipulations. Numbers in parentheses are the detection thresholds, while Detect is the portion of} \\
\!\!\!\!\emph{the images meeting them. For attacks with multiple configurations, lower rows correspond to stronger distortions. $*$ denotes overly-destructive configurations.}
\vspace{2mm}
\end{tabular}

\setlength{\tabcolsep}{0.5pt}
\centering
\resizebox{\textwidth}{2.3cm}{
\fontsize{13pt}{9pt}\selectfont
\begin{tabular}{llll"cc|cc|cc|cc|cc|cc|cc}
\thickhline
\addlinespace[5pt]
& & & & \multicolumn{2}{c}{\textit{\textbf{Yu1}} (61\%)} & \multicolumn{2}{c}{\textit{\textbf{Yu2}} (63\%)} & \multicolumn{2}{c}{\textit{\textbf{HiDDeN}} (73\%)} & \multicolumn{2}{c}{\textit{\textbf{PTW}} (70\%)} & \multicolumn{2}{c}{\textit{\textbf{StableSig.}} (69\%)} & \multicolumn{2}{c}{\textit{\textbf{StegaStamp}} (63\%)} & \multicolumn{2}{c}{\textit{\textbf{TRW}} (0.0141)}\\
\multicolumn{4}{c"}
{\diaghead(-4,1){aaaaaaaaaaaaaaaaaaaaaa}%
{{\fontsize{13pt}{11pt}\selectfont \textit{Manipulation}}}{{\fontsize{13pt}{9pt}\selectfont \textit{Scheme}}}} & Bit Acc$\uparrow$ & Detect$\uparrow$ & Bit Acc$\uparrow$ & Detect$\uparrow$ & Bit Acc$\uparrow$ & Detect$\uparrow$ & Bit Acc$\uparrow$ & Detect$\uparrow$ & Bit Acc$\uparrow$ & Detect$\uparrow$ & Bit Acc$\uparrow$ & Detect$\uparrow$ & Inv. Dist.$\uparrow$ & Detect$\uparrow$\\ 
\midrule
\textbf{\textit{None}} & & & & 98.42\% & 100\% & 99.71\% & 100\% & 99.16\% & 100\% & 97.96\% & 100\% & 98.97\% & 100\% & 99.95\% & 100\% & 0.022 & 100\% \\ \thickhline
\addlinespace[2pt]
\textbf{\textit{Cropping}} & \multicolumn{3}{c"}{\Large \textit{ratio = 10\%}} & 74.56\% & 100\% & 53.71\% & 0\% & 86.84\% & 100\% & 93.9\% & 100\% & 97.86\% & 100\% & 93.96\% & 100\% & 0.0155 & 100\% \\ \thickhline
\addlinespace[2pt]
\multirow{2}{*}{\bf \textbf{\textit{Blur}}} & \multicolumn{3}{c"}{\Large \textit{std. = 0.5}} & 98.13\% & 100\% & 99.55\% & 100\% & 96.45\% & 100\% & 96.93\% & 100\% & 98.52\% & 100\% & 99.95\% & 100\% & 0.0221 & 100\% \\
& \multicolumn{3}{c"}{\Large \textit{std. = 1.0}} & 94.6\% & 100\% & 99.73\% & 100\% & 62.15\% & 5.00\% & 82.25\% & 100\% & 86.65\% & 99\% & 99.88\% & 100\% & 0.0207 & 100\% \\ \thickhline
\addlinespace[2pt]
\multirow{3}{*}{\bf \textbf{\textit{Guided Blur}}} & \multicolumn{3}{c"}{\Large \textit{color std. = 0.1}} & 87.91\% & 100\% & 99.32\% & 100\% & 75.6\% & 76\% & 94.41\% & 100\% & 96.38\% & 100\% & 99.91\% & 100\% & 0.0212 & 100\% \\
& \multicolumn{3}{c"}{\Large \textit{color std. = $1.0^*$}} & 85.32\% & 100\% & 99.43\% & 100\% & 56.12\% & 0\% & 78.66\% & 98\% & 71.25\% & 69\% & 99.93\% & 100\% & 0.02 & 100\% \\
& \multicolumn{3}{c"}{\Large \textit{color std. = $5.0^*$}} & 84.9\% & 100\% & 99.62\% & 100\% & 53.81\% & 0\% & 76.42\% & 92\% & 66.54\% & 50\% & 99.93\% & 100\% & 0.0199 & 100\% \\ \thickhline
\addlinespace[2pt]
\multirow{2}{*}{\bf \textbf{\textit{Quantization}}} & \multicolumn{3}{c"}{\Large \textit{factor = 8}} & 85.86\% & 100\% & 99.16\% & 100\% & 89.34\% & 98\% & 74.92\% & 93\% & 98.83\% & 100\% & 99.83\% & 100\% & 0.0207 & 100\% \\
& \multicolumn{3}{c"}{\Large \textit{factor = 10}} & 82.51\% & 100\% & 98.7\% & 100\% & 86.51\% & 98\% & 68.34\% & 46\% & 99.04\% & 100\% & 99.8\% & 100\% & 0.0201 & 100\% \\ \thickhline
\addlinespace[2pt]
\multirow{2}{*}{\bf \textbf{\textit{JPEG}}} & \multicolumn{3}{c"}{\Large \textit{quality = 110}} & 89.58\% & 100\% & 99.55\% & 100\% & 70.73\% & 43\% & 87.66\% & 100\% & 97.68\% & 99\% & 99.94\% & 100\% & 0.0215 & 100\% \\
& \multicolumn{3}{c"}{\Large \textit{quality = 80}} & 76.38\% & 100\% & 93.15\% & 100\% & 66.48\% & 16\% & 72.17\% & 76\% & 90.66\% & 100\% & 99.89\% & 100\% & 0.202 & 100\% \\
\thickhline \\
\end{tabular}
}
\end{table*}

By monitoring the spectral loss, we can determine when it stops increasing for each value of $c_{{\ell_p}}$ and the number of steps after which updating $c_{{\ell_p}}$ does not result in improvements. This yields $5000$ iterations for images smaller or larger than {\small \mbox{$256 {\times} 256$}}, and $2000$  otherwise. The number of binary steps is $2$ for sizes $\leq$ {\small \mbox{$256 {\times} 256$}} and $4$ otherwise. The optimizer is \textit{Adam} with learning rate $2 \cdot 10^{-4}$, reduced by $0.5$ when no improvement (above $10^{-4}$) occurs for $10$ iterations. We clip gradients at $5 \cdot 10^{-3}$ for \textit{DVL} and $5 \cdot 10^{-2}$ for \textit{Alex} to ensure they do not dominate the process.

\noindent
\textit{\underline{Low-frequency Disruptions} (Stage 2---~\S\ref{subsec:semantic})}: Each filter is characterised by the dimensionality \mbox{\small$M_t {\times} N_t$} of ${\mathcal{K}^*\!\!\!_{_t}}$ and permissiveness $\!\!{\sigma\!_{c_{_t}}}$. We find ${\sigma\!_{c_{_t}}}\!\!=\!\!5 \cdot 10^{-2}$ optimal. When choosing the kernel sizes and number of filters $T$, resources may be prohibitive as we have a kernel per pixel, making a filter scale as the number of pixels multiplied by its kernel size. We found $T\!\!\!=\!\!9$ with different sizes defeats all studied systems. This architecture was chosen (semi-randomly) to target various potential geometries of the watermark under our computational limitations (\textit{$40$GB} \textit{A100 GPU}) but can be extended with better hardware. Yet, our findings indicate that this may not be necessary. The ordered sizes are $(21,5),\!$ $(5,5), (17,33), (7,7), (47,5), (33,17), (17,17), (5,5)\!\!\!$ and $(3,3)$. Our spectral loss consists of \textit{MPL} and \textit{FRL}. While we maximize both, \textit{FRL} is a supporting regularization term--- see~\S\ref{subsubsec:low_freq_loss}. Hence, we deem a sample optimal if it maximizes \textit{MPL}. Yet, $\lambda_{{\text{\tiny \mbox{$FRL$}}}}$ must account for \textit{FRL}, ensuring the outputs meet the conditions it enforces. $\lambda_{{\text{\tiny \mbox{$FRL$}}}}\!\!=\!\!5$ was found optimal, but it can be adjusted if the two components do not meet expectations (see~\S\ref{subsubsec:low_freq_loss}). We found $\lambda_{{\|\|_2}}\!\!=\!\!2.5\cdot 10^{-4}$ ideal but it can be calibrated similarly.

We universally use \textit{DVL} (see~\S\ref{subsec:high_freq}). As low-frequency perturbations do not introduce sporadic peaks, the remaining parameters can be selected uniformly with little regard to the image size. The exception is the threshold $t_{{{\text{\tiny \mbox{${DVL}$}}}}}$, as smaller images ({\small \mbox{$\leq 256 {\times} 256$}}) are more sensitive to structural changes (unlike high-frequency perturbations that these images' low resolution masks), exhibiting visible differences for $t_{{{\text{\tiny \mbox{${DVL}$}}}}}>10^{-3}$, while larger images can withstand $3\times$ this change. With an initial $c_{{{\text{\tiny \mbox{${DVL}$}}}}}=10^6$, a single step running for $500$ iterations is sufficient. We use \textit{Adam} with a $10^{-2}$ learning rate for the filters and $2 \cdot 10^{-3}$ for $\delta$, decayed by $0.99$ if \textit{MPL} fails improve by $5 \cdot 10^{-2}$ for $10$ iterations. As \textit{DVL}'s impact is weakened due to this procedure's structural nature, we allow its gradients to be more dominant, clipped at $1.0$.

Filters operate on all image channels identically. Hence, as the reason behind including the modifier $\delta$ is the filters' ability to smoothen its otherwise high-frequency changes, $\delta$ must also affect all channels similarly. Thus, we use a \textit{gray-scale} modifier $\delta$ that applies the same changes to all channels. While manipulating spectral amplitudes impacts the texture and results in less critical changes, phase modifications are more visible (see~\S\ref{subsubsec:candidates}). Hence, we should ideally restrict our filters to manipulate magnitudes. The frequency response of a filter does not change the phase \textit{iff} it is even-symmetric (identical to the reflections of each of its quadrants across both axes). While this holds for \textit{basic} (convolutional) filters, ours will converge to similar kernels for neighboring pixels that are of equal visual importance and can be altered similarly, indicating that each region will behave as if it were ``convolved with a basic kernel". Thus, our filters should possess this symmetry. Accordingly, instead of optimizing all weights, we focus only on the first quadrant of each kernel and reflect it across both axes.

\section{Watermarking Schemes' Configuration}
\label{app:wat_len}
The (author-reported) watermark lengths are in Table~\ref{tab:wat_len}.

\begin{table}[h!]
    \centering
    \caption{Watermark lengths.}
    \label{tab:wat_len}
\fontsize{8pt}{7pt}\selectfont
\begin{tabular}{l}
\\
\!\!\!\!\emph{\underline{Note:} \textit{TRW}'s watermarks are not binary; thus, this metric is not applicable.}
\vspace{2mm}
\end{tabular}
    \setlength{\tabcolsep}{0.5pt}
    \begin{tabular}{|c|c|}
    \thickhline
        Scheme & Length (\textbf{\# bits}) \\ \thickhline
        \textit{Yu1} & 128 \\ \hline
        \textit{Yu2} & 100 \\ \hline
        \textit{HiDDeN} & 30 \\ \hline
        \textit{PTW} & 40 \\ \hline
        \textit{StableSignature} & 48 \\ \hline
        \textit{StegaStamp} & 100 \\ \hline
        \textit{TRW} & NA \\ \thickhline 
    \end{tabular}
\end{table}

\section{Baseline Results Against Image Manipulations}
\label{app:baselines}

 \begin{figure*}[t!]
  \centering
\includegraphics[width=\linewidth,height=11.5cm]{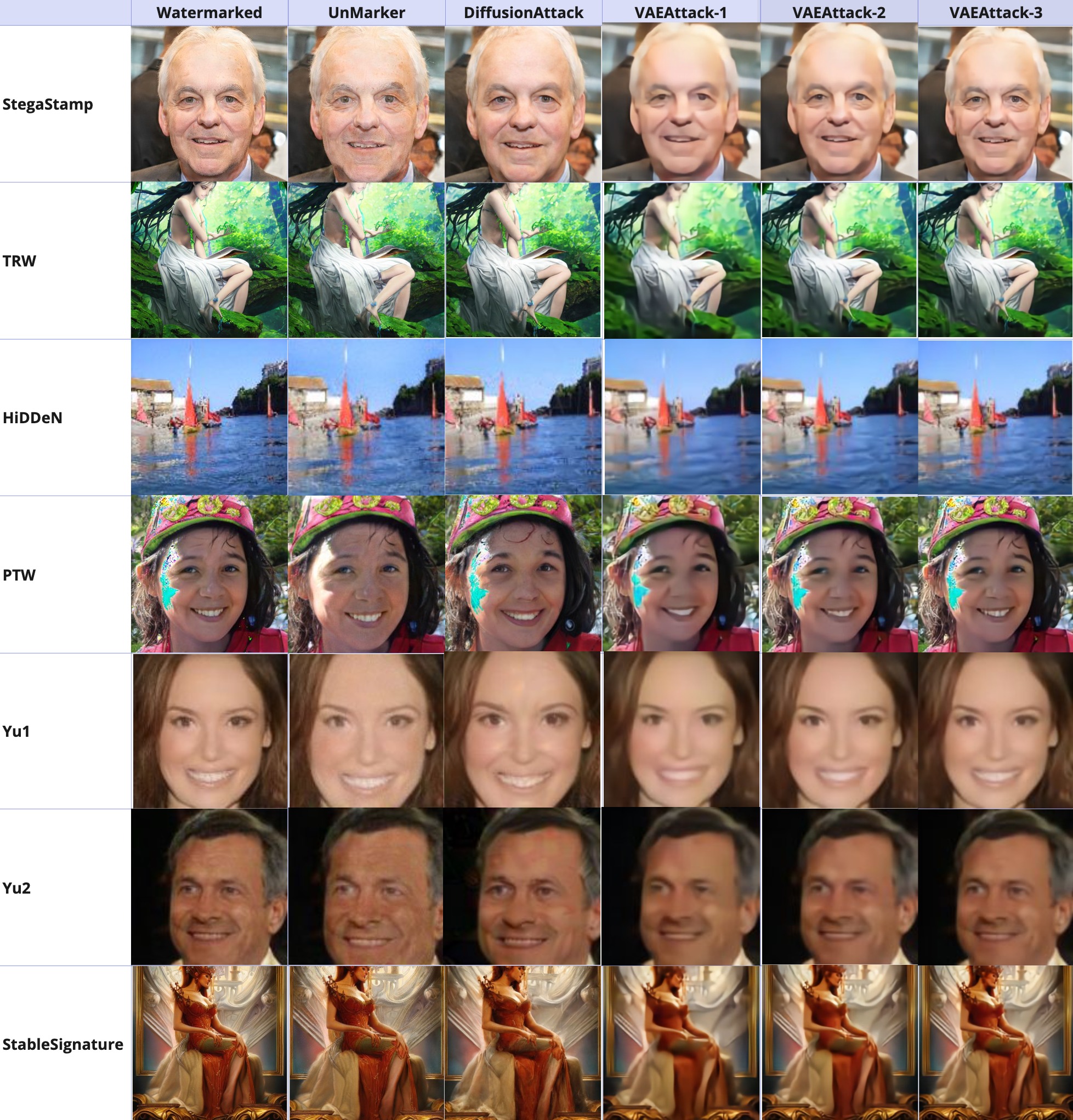}
  \caption{Outputs of the three removal attacks. The \textit{VAEAttack} over-smooths images, resulting in them losing crucial information and occasionally looking cartoonish. Compared to the \textit{DiffusionAttack}, \textit{UnMarker}'s outputs better resemble the watermarked images. The \textit{DiffusionAttack} also introduces semantic incoherences such as additional teeth in the case of \textit{PTW} and omits identifying details such as freckles on the forehead for \textit{StegaStamp}. \textit{PTW}'s watermarked image itself is of inferior quality with an unnatural patch. The \textit{DiffusionAttack} retains this artifact while \textit{UnMarker} eliminates it, enhancing the image.}
  \label{fig:comp}
\end{figure*}

\noindent 
The kernel size for \textit{Gaussian (Guided) Blurring} is {\small \mbox{$5 {\times} 5$}}. We exploit \textit{Guided Blurring}'s edge-preserving properties to increase the \textit{std.} (to $5.0$). To control the modulation at edges, these accept an additional \textit{color kernel}'s \textit{std.} (see~\S\ref{subsec:semantic}), for which we have three values: $0.1,\; 1.0$, and $5.0$. Yet, $1.0$ and $5.0$ are overly destructive and do not preserve quality, chosen only to explore robustness. All parameters are in Table~\ref{tab:manipulations}.

\noindent
\textit{\underline{Results:}} Most systems are robust to all \textit{practical} manipulations. Despite resisting all other modifications, \textit{Yu2} breaks completely under \textit{cropping}. Hence, unlike other schemes, it highly depends on spatial attributes\footnote{Yu et al.~\cite{yu2} only evaluate a single \textit{ProGAN}~\cite{progan} architecture under manipulations, whereas we chose their \textit{StyleGAN2}~\cite{stylegan2} model.}. \textit{Yu1, HiDDeN} and \textit{TRW} exhibit a considerable drop in the \textit{bit accuracies} (\textit{inverse distance}) after \textit{cropping}, although their overall detection is not affected. This is not surprising, as the watermarking spectral modifications must be spread throughout the image to minimize visual cues, implying that a portion may be allocated to the background. Yet, they retain robustness by restricting most of these changes to central regions.

\textit{Yu1, StegaStamp} and \textit{TRW} resist all manipulations, while \textit{StableSignature} can withstand realistic, non-obtrusive modifications, making it also robust for practical purposes. \textit{PTW} is slightly less powerful, losing some of its efficacy under even non-destructive operations, such as \textit{Quantization} with a factor of $10$ and \textit{JPEG} with quality $80$ that decrease its performance to $46\%$ and $76\%$, respectively. \textit{HiDDeN} exhibits similar behavior, resisting all realistic manipulations aside from \textit{JPEG} that brings its detection to $43\%$ even when used with a reasonably good quality of $110$ or \textit{Blurring} with $std.=1.0$ that lowers its detection score to $5.0\%$. 

It is expected that some \textit{non-semantic} systems (e.g., \textit{PTW, HiDDeN} and \textit{StableSignature}) begin to break under spectral manipulations such as \textit{JPEG} and \textit{Blurring}. As explained in~\S\ref{subsection:freq}, constructing spectral \textit{non-semantic} watermarks should be done in an input-specific manner that is also persistent, which is a challenging task that will be restricted to volatile watermarks unless the scheme is sufficiently advanced. This explains why the more sophisticated \textit{StableSignature} and \textit{PTW} outperform \textit{HiDDeN}. Yet, allowing for severe manipulations via stronger \textit{(Guided) Blurring} breaks \textit{StableSignature} as well, although this is only included to demonstrate this phenomenon that attests to the spectral nature of these watermarks and is not a valid attack (see~\S\ref{subsec:setup}). \textit{Quantization} can cause sharp pixel shifts (high-frequency noise), explaining \textit{PTW}'s vulnerability.

With few exceptions, which we attribute to a slight mismatch in the experimental settings, our findings align with those attained by the authors of the schemes that evaluate the same attacks with similar parameters.


\section{Effects of Adversarial Filtering}
\label{app:filt}
\begin{table}[h]
\caption{\textit{UnMarker}'s performance against \textit{semantic} schemes with and without optimizable filters.}
\label{tab:filters}
\renewcommand{\arraystretch}{7.5}
\fontsize{210pt}{30pt}\selectfont
\setlength{\tabcolsep}{0.5pt}
\resizebox{\columnwidth}{0.8cm}{
\begin{tabular}{ll@{\hspace{-5ex}}"@{\hspace{1ex}}cccc|@{\hspace{1ex}}cccc}
\topthickhline
\addlinespace[10pt]
& & \multicolumn{4}{c}{\textit{\textbf{StegaStamp}} (63\%)} & \multicolumn{4}{c}{\textit{\textbf{TRW}} (0.0141)}\\ 
\multicolumn{2}{l"@{\hspace{1ex}}}
{\diaghead(-6,1){aaaaaaaaaaaaaaaaaaaaaaaa}%
{\fontsize{250pt}{9pt}\selectfont \textit{Attack}}{\fontsize{250pt}{9pt}\selectfont \textit{Scheme}}} & \textit{FID}$\downarrow$ & \textit{LPIPS}$\downarrow$ & Bit Acc$\downarrow$ & Detect$\downarrow$ & \textit{FID}$\downarrow$ & \textit{LPIPS}$\downarrow$ & Inv. Dist.$\downarrow$ & Detect$\downarrow$ \\ [-5pt]
\midrule
\multirow{2}{*}{{\fontsize{250pt}{10pt}\selectfont
\textbf{\textit{UnMarker}}}} & \;\;{\fontsize{230}{40}\selectfont \textit{with filters}} & 34.69 & 0.08 & 61.49\% & 43\% & 60.73 & 0.1 & 0.014 & 40\% \\
& \;\;{\fontsize{230}{40}\selectfont \textit{w/o filters}} & 30.45 & 0.04 & 93.13\% & 100\% & 49.59 & 0.04 & 0.0147 & 86\% \\ 
\thickhline
\end{tabular}
}
\end{table}
The reader may wonder whether the low-frequency loss alone can defeat \textit{semantic} schemes, eliminating the need for our filters. Based on our arguments explaining how direct optimizations are of limited applicability against low-frequency watermarks due to their non-systemic nature (see~\S\ref{subsec:semantic}), this is not the case. For validation, we compare \textit{UnMarker} against \textit{semantic} schemes with and without our filters. In both experiments, the loss is the same from~\S\ref{subsubsec:low_freq_loss}. The results in Table~\ref{tab:filters} prove the filters necessary. While \textit{TRW} experiences a slight degradation without filters, including them is far more effective, bringing its detection to $40\%$ (compared to $86\%$ otherwise). \textit{StegaStamp} is unaffected when filters are excluded, retaining $100\%$ detection. With filters, this drops to $43\%$. We also verified the filter-free attack's ineffectiveness persists even if the perturbation slightly increases. Although similarity is slightly worsened by our filters as they encourage structural changes, the scores show that, even with filtering, \textit{UnMarker} retains quality.

\section{Outputs of \textit{UnMarker} vs. The \textit{SOTA}}
\label{app:fig_comp}
Fig.~\ref{fig:comp} displays examples of watermarked images and the outputs obtained by subjecting them to advanced attacks.

%% file: Sections/meta_review.tex
\newpage 

\section{Meta-Review}

The following meta-review was prepared by the program committee for the 2025 IEEE Symposium on Security and Privacy (S\&P) as part of the review process as detailed in the call for papers.

\subsection{Summary}
This paper introduces \textit{UnMarker}, a universal attack method designed to compromise defensive image watermarking mechanisms.

\subsection{Scientific Contributions}
\begin{itemize}
\item Identifies an Impactful Vulnerability
\item Provides a Valuable Step Forward in an Established Field
\item Independent Confirmation of Important Results with Limited Prior Research
\end{itemize}

\subsection{Reasons for Acceptance}
\begin{enumerate}
\item \textit{UnMarker} demonstrates effectiveness against multiple watermarking schemes.
\item The paper provides valuable insights into the vulnerabilities of current watermarking methods by leveraging spectral amplitude analysis.
\end{enumerate}
